\documentclass[aps,twocolumn,amsmath,amssymb,eqsecnum,prb]{revtex4-1}
\usepackage{graphicx}
\usepackage[colorlinks=false]{hyperref}

\newcommand {\be}{\begin{equation}}
\newcommand {\ee} {\end{equation}}
\newcommand {\bea}{\begin{eqnarray}}
\newcommand {\eea} {\end{eqnarray}}
\newcommand{\non}{\nonumber}

\newcommand{\tr}{\mathrm{Tr}}
\newcommand{\ssc}{{\alpha}}

\newcommand{\ssh}{{\beta}}
\newcommand{\sss}{{\gamma}}
\newcommand{\gammat}{{\widetilde{\gamma}}}
\newcommand{\Tt}{{\widetilde{T}}}
\begin{document}
\title{Low-energy effective theory in the bulk for transport in a topological phase}
\author{Barry Bradlyn}
\email{barry.bradlyn@yale.edu}
\author{N. Read}
\affiliation{Department of Physics, Yale University, P.O. Box 208120, New Haven, Connecticut 06520-8120,
USA}
\date{February 15, 2015}
\begin{abstract}
We construct a low-energy effective action for a two-dimensional non-relativistic topological (i.e.\
gapped) phase of matter in a continuum, which completely describes all of its bulk electrical, thermal,
and stress-related properties in the limit of low frequencies, long distances, and zero temperature,
without assuming either Lorentz or Galilean invariance.
This is done by generalizing Luttinger's approach to thermoelectric phenomena, via the introduction of a
background vielbein (i.e.\ gravitational) field and spin connection a la Cartan, in addition to the
electromagnetic vector potential, in the action for the microscopic degrees of freedom (the matter fields).
Crucially, the geometry of spacetime is allowed to have timelike and spacelike torsion.
These background fields make all natural invariances--- under U(1) gauge transformations,
translations in both space and time, and spatial rotations---appear locally, and corresponding conserved
currents and the stress tensor can be obtained, which obey natural continuity equations. On integrating
out the matter fields, we derive the most general form of a local bulk induced action to first order
in derivatives of the background fields, from which thermodynamic and transport properties can be
obtained. We show that the gapped bulk cannot contribute to low-temperature thermoelectric transport
other than the ordinary Hall conductivity; the other thermoelectric effects (if they occur) are thus
purely edge effects. The coupling to ``reduced'' spacelike torsion is found to be absent in
minimally-coupled models, and using a generalized Belinfante stress tensor, the stress response to
time-dependent vielbeins (i.e.\ strains) is the Hall viscosity, which is robust against perturbations
and related to the spin current as in earlier work.
\end{abstract}

\maketitle

\section{Introduction}\label{sec1}
There has been great interest recently in the thermal Hall conductivity of gapped topological phases at
temperatures small compared with the bulk energy gap. It
has been known for some time, using arguments based on the existence of a gapless edge, that the thermal
Hall conductivity $\kappa^H$ of such systems is given by\cite{Kane1997,Read2000}
\begin{equation}
\kappa^H=\frac{\pi T}{6}c, \label{thermalhallresult}
\end{equation}
where $T$ is the temperature and $c=c_L-c_R$ is the (topological) central charge of the edge theory. It
has been hoped that a calculation of $\kappa^H$ could be carried out which would illustrate the appearance
of the central charge from bulk correlation functions. On the other hand, it has been
pointed out\cite{Read2000} that the central charge appears as the coefficient of the gravitational
Chern-Simons term
\begin{equation}
S_\mathrm{GCS}=\frac{c}{96\pi}\int d^3x\, \widehat{\epsilon}^{\mu\nu\lambda}\left(\Gamma^\rho_{\mu\sigma}
\partial_{\nu}\Gamma^{\sigma}_{\nu\rho}+\frac{2}{3}\Gamma^{\rho}_{\mu\sigma}
\Gamma^{\sigma}_{\nu\theta}\Gamma^{\theta}_{\lambda\rho}\right),
\end{equation}
in terms of the Christoffel symbols $\Gamma_{\mu\nu}^\lambda$. This term, however, is of too high an
order in derivatives of the metric to
describe thermal conductivity directly\cite{Stone2012}, but is nonetheless connected with the central
charge of the edge states. It manifests itself in the bulk rather through the response
of the energy-momentum-stress tensor to gradients in curvature.

In this paper we show that the long-wavelength \emph{bulk} thermal transport properties are
completely independent of the central charge, and that the only non-vanishing bulk thermoelectric
current is the ordinary Hall current (we neglect effects that vanish exponentially in the energy gap
over temperature as the temperature goes to zero). We do this by constructing the most general low-energy
effective action for the bulk at the correct order in derivatives of external vielbein
(i.e.\ gravitational) and electromagnetic fields. We show
that responses to certain gradients of the vielbeins correspond to thermal and thermoelectric response
functions. After using general thermodynamic arguments to identify terms in the effective action, we
show that the bulk thermal currents in response to gradients of the vielbeins yield purely
\emph{magnetization} currents, that is currents that vanish when integrated along any cross-section of
the sample. Thus, we show that the bulk contribution to the thermal Hall conductivity is exponentially
suppressed due to the gap.

Our formalism treats arbitrary background geometries for a non-relativistic system that has neither
Lorentz nor Galilean invariance. This unified approach allows us also to consider the stress response
to background fields, and thus viscosity, on the same footing as the thermoelectric effects, and to account
for all bulk magnetization effects: number, energy, and momentum magnetizations. When we use the formalism
to study Hall viscosity of a topological phase, we find that use of the appropriate Belinfante stress
tensor, while not affecting the results for thermoelectric coefficients, has the effect of removing
the contribution of spacelike torsion to the Hall viscosity that had been found in a relativistic
setting\cite{Hughes2011,Hughes2013}. We also point out that for simple non-relativistic models in which
the background
fields are minimally coupled, there is no coupling to spacelike torsion in the limit of a trivial
spacetime without torsion. Instead, the Hall viscosity and the spin current
follow\cite{Read2009b,Hoyos2012} purely from the Wen-Zee term\cite{Wen1992} in the effective action,
and are related as in previous work\cite{Read2009,Read2011}.

Our analysis of thermoelectric transport will make contact with the formalism developed by Cooper, Halperin
and Ruzin (CHR) \cite{Cooper1997}, so we will now recapitulate their main points. They consider the number
current $\mathbf{J}$ and  $\mathbf{J}_{\mathrm{E}}$ in the presence of an applied electric field
$\mathbf{E}=-\nabla\phi$ and a fictitious gravitational field $\psi$ coupled to the energy
density\cite{Luttinger1964}. By considering linear response to these fields in the bulk via the Kubo
formula, they obtain a set of zero frequency and zero wave vector response functions $L^{(n)}_{ij}$,
such that the changes in number and energy current density can be expressed as
\begin{align}
\delta J^i&=-L^{(1)}_{ij}\partial_j\phi-L^{(2)}_{ij}\partial_j\psi ,\nonumber\\
\delta J_\mathrm{E}^i&=-L^{(3)}_{ij}\partial_j\phi-L^{(4)}_{ij}\partial_j\psi.
\end{align}
(Here $i$, $j=1$, $2$ are the space coordinate indices, the summation convention is in effect for these
indices, and we will also use $\epsilon_{ij}=-\epsilon_{ji}$, with $\epsilon_{12}=1$.)
They note, however, that in the presence of a background magnetic field $B$ (perpendicular to the plane)
there exist magnetization number and energy currents, which if the bulk is translation
invariant appear only as edge currents. These are given (in our notation for two space dimensions) by
\begin{align}
J_{\mathrm{mag}}^i&=\epsilon_{ij}\partial_jm ,\nonumber\\
J_{\mathrm{E,mag}}^i&=\epsilon_{ij}\partial_jm^\mathrm{E}, \label{magcurrents}
\end{align}
where $m$ is the ordinary magnetization density and $m^\mathrm{E}$ is a suitably defined ``energy
magnetization'' density. In the presence of the fields $\phi$ and $\psi$ the magnetizations differ from
their unperturbed values $m_0$ and $m^{\mathrm{E}}_0$ by
\begin{align}
m&=\left(1+\psi\right)m_0 ,\nonumber\\
m^\mathrm{E}&=\left(1+2\psi\right)m_0^\mathrm{E}+\phi m_0. \label{magperts}
\end{align}
The magnetization currents induced by the external fields must be accounted for in order to obtain
the transport current densities $\mathbf{J}_\mathrm{tr}$ and $\mathbf{J}_{\mathrm{tr,E}}$; the
transport current densities by definition give the net current across a section when integrated along it,
and are defined to occur solely in the bulk. They find
\begin{align}
J^i_\mathrm{tr}&=-L^{(1)}_{ij}\partial_j\phi-\left(L^{(2)}_{ij}+m_0\epsilon_{ij}\right)\partial_j
\psi ,\\
J_\mathrm{E,tr}^i&=-\left(L^{(3)}_{ij}+m_0\epsilon_{ij}\right)\partial_j\phi-\left(L^{(4)}_{ij}
+2m^\mathrm{E}_0\epsilon_{ij}\right)\partial_j\psi.
\end{align}
Then CHR used generalized Einstein relations, that with chemical potential $\mu$ and non-zero temperature
$T$ (both of which can be position dependent since the system is not in equilibrium) the transport currents
are responses only to the combinations
$\nabla\psi+(1/T)\nabla T$ and $\nabla\phi+T\nabla(\mu/T)$. Finally setting $\psi=0$, and defining
$\xi=\phi+\mu$ and the transport heat current density
$\mathbf{J}_\mathrm{Q,tr}=\mathbf{J}_E-\xi\mathbf{J}_\mathrm{tr}$, CHR showed that
\begin{align}
J^i_\mathrm{tr}&=-N^{(1)}_{ij}\partial_j\xi-\frac{1}{T}N^{(2)}_{ij}\partial_jT, \\
J_\mathrm{Q,tr}^i&=-N^{(3)}_{ij}\partial_j\xi-\frac{1}{T}N^{(4)}_{ij}\partial_jT, \label{transresp}
\end{align}
with
\begin{align}
N^{(1)}_{ij}=&L^{(1)}_{ij}, \label{transportresp1}\\
N^{(2)}_{ij}=&L^{(2)}_{ij}-\mu L^{(1)}_{ij}+m_0\epsilon_{ij} ,\\
N^{(3)}_{ij}=&L^{(3)}_{ij}-\mu L^{(1)}_{ij}+m_0\epsilon_{ij}, \\
N^{(4)}_{ij}=&L^{(4)}-\mu\left(N^{(2)}_{ij}+N^{(3)}_{ij}\right)-\mu^2L^{(1)}_{ij}+2m^\mathrm{E}_0
\epsilon_{ij}. \label{transportresp2}
\end{align}
Here the coefficient matrices $N$ obey the Onsager relations, for example
that $N^{(2)}_{ij}(B)= N^{(3)}_{ji}(-B)$, (as do the matrices $L$) whereas the local current responses
to $\nabla\mu$ and $\nabla T$ do not.
We see also that $N^{(2)}_{ij}$ and $N^{(4)}_{ij}$ must vanish
faster than $T$ as $T\to0$, because the corresponding conductivities must vanish in that limit.
In what follows, we will show how the bulk contributions to these coefficients appear in and can be
determined from the low-energy effective action for the bulk of a gapped system.
It will follow that for such gapped systems, among these coefficients only
$N^{(1)}$ (the Hall conductivity) receives a bulk contribution. The appearance of the central charge
in $N^{(4)}$ is due to an edge effect.

We should explain the type of systems to which our formalism applies. We assume that the system is gapped
in the bulk, so when we integrate out the matter fields only integrals of local expressions can occur in
this ``induced'' action in the bulk. We assume this action depends only on the background electromagnetic
field and on the spacetime geometry, and that it has symmetries under U($1$) gauge transformations (because
particle number is conserved), coordinate transformations (from translation invariance in both space and
time in a flat background, leading to conservation of energy and momentum), and spatial rotations. Thus
we assume that these symmetries are not broken either spontaneously or explicitly. If either occurred, it
would be necessary to include further background fields in the induced action that describe the breaking,
and for spontaneously broken continuous symmetry in a system with short range interactions there would be
gapless degrees of freedom, so that the induced action is not local. Hence our approach applies to quantum
Hall systems and to insulators (including topological insulators) in a continuum approximation with
rotation invariance, but not to fluids or (possibly
topological) superconductors. In a model for a superconductor in which particle number is conserved,
either (in the case of short-range interactions) it has a gapless Goldstone mode, or with a long-range
interaction it can be fully gapped, but then the long range of interaction produces additional problems
for us. Without conserved particle number we could simply drop the U($1$) gauge field everywhere, but
such paired states of fermions in which the pairs have non-zero angular momentum also break rotation
symmetry and require a different treatment that will not be given here.

In Section \ref{sec2}, we explain the geometry to be used, and develop our microscopic model for the
deformed system deriving explicit expressions and conservation laws for the currents. After discussing
in Sec.\ \ref{sec:gen} some general facts about different terms in an induced action, we then write
down in Section \ref{sec3} the most general effective (induced) action to linear order in derivatives of
the perturbing fields and consistent with the symmetries of the microscopic model. These results allow us
to identify number, energy, and momentum magnetizations. Then in Section
\ref{respsec} we turn to linear response. In Sec.\ \ref{sec:thermo} we calculate the response of the
number and heat currents to an electric field and to Luttinger's gravitational field, and show explicitly
that the bulk contributions to the thermoelectric transport currents vanish. In Sec.\ \ref{sec:stress},
we address the stress response. We go over to a generalized Belinfante definition of the
energy-momentum-stress tensor, which is described in Appendix \ref{appB}. This eliminates contributions
to the Hall viscosity from locally-invariant terms in the bulk (the non-relativistic version of the
``torsional Hall viscosity''\cite{Hughes2013} is one such effect). We observe that in simple models,
there is no momentum magnetization. Finally we show that the Wen-Zee term produces the Hall viscosity in
agreement with the spin current, in line with previous results.

There seems to be some confusion in the literature about whether the thermal Hall conductivity comes
from the bulk. For free fermion systems, one can use linear response theory to derive the thermal
conductivity\cite{Smrcka1977,Streda1983,Sumiyoshi2012,Qin2011} in a way that seemingly makes no reference
to the edge physics. In Appendix \ref{app} we review the calculation of the thermoelectric response
coefficients for a non-interacting integer quantum Hall system. The key point is that such approaches
calculate the response of current densities integrated across sections of a sample. Such integrated
currents implicitly contain contributions from edge physics. The
calculation thus essentially reduces to the use of the same edge argument to which we already
referred\cite{Kane1997,Read2000}.

\section{Background fields and matter action}\label{sec2}
\subsection{Spacetime geometry}

Before we begin, let us establish some notational conventions. We will be working in $d+1$ spacetime
dimensions throughout, with coordinates $x^\mu$. We will need to distinguish between two different types
of indices: ambient spacetime indices, denoted by  $\mu$, $\nu=0$, $1$, $2$, \ldots $d$, and similar
indices, denoted by $\alpha$, $\beta=0$, $1$, $2$, \ldots, $d$ in a flat internal spacetime. When we refer
to space-like directions only, we will use Roman letters, $i$, $j=1$, $2$, \ldots for the ambient indices,
and $a$, $b=1$, $2$, \ldots for the internal indices. We use the summation convention for all four types
of indices, adhering from this point on to the conventions of placement of upper and lower indices,
and $\partial_\mu=\partial/\partial x^\mu$.
In this initial discussion, we keep $d$ general, but later we specialize to
$d=2$.

The geometry of the spacetime that we use does not possess the metric structure of Minkowski spacetime, or
even of a Galilean analog. The only structure is that at any point we can distinguish
between space and time, as if there were a local absolute time coordinate, and a local positive-definite
spatial metric. These statements do {\em not} mean that an absolute time coordinate can be defined, even on
a small region of spacetime, so neither are there spacelike surfaces of fixed time. (These structures
are similar to those used by Wen and Zee\cite{Wen1992}, however they assumed that the spacetime has a
global absolute time, and that the spatial metric of fixed-time slices was time independent.)
In order to precisely define these structures, we prefer to be able to use arbitrary coordinate systems,
and to be able to make arbitrary coordinate transformations (diffeomorphisms). The spacetime structures can
be introduced using Cartan's vielbein formalism \cite{Carroll2004} (also called the vierbein or tetrad
formalism in the case of $d=3$, or $3+1$ dimensions). At each point we have a set of one-forms
with components $e_\mu^\alpha$ and a dual set of vector fields $e_\alpha^\mu$, that obey the duality
relations $e^\mu_\alpha e_\mu^\beta=\delta_\alpha^\beta$ and $e^\mu_\alpha e_\nu^\alpha=\delta_\mu^\nu$.
Either set defines a frame at each point in spacetime, that is a preferred basis set of one-forms (or the
dual set of vectors) indexed by $\alpha$; these define the structure in a coordinate-independent way. The
frames are assumed not to degenerate at any spacetime point; that is, the set of vectors is linearly
independent at each $x$.
Actually, the choice of basis (in the internal space) for the spacelike one-forms $e_\mu^a$ (or for
the spacelike vectors $e^\mu_a$) is arbitrary up to a rotation on the internal indices; we incorporate that
fact in due course.
The vielbeins and their inverses can be used (by contraction) to convert ambient to internal spacetime
indices or vice versa.

In particular, we have a one-form with components $e_\mu^0$, where the upper index is internal.
If there existed a function $t$ of position over regions of spacetime, such that (using the notation
of differential forms) $e_\mu^0 dx^\mu=dt$, then $t$ would be absolute time, but we do {\em not}
assume this. In order to obtain such an absolute time $t$, the necessary and sufficient condition
is $\partial_\nu e_\mu^0-\partial_\mu e_\nu^0=0$; in general we do {\em not} impose this. We can use
the one-form to measure amounts of time using a squared line element for the time direction (or a
particular degenerate or ``partial''  metric)
\be
(e_\mu^0 dx^\mu)^2.
\ee

Likewise for the analog of time slices, we have the components $e_\mu^a$ and $e^\mu_a$, and the internal
spacelike components of each are orthogonal to the time-like component of the inverses, for example
$e_a^\mu e_\mu^0=0$, just as if the vectors $e^\mu_a$ were tangent vectors to a fixed-time surface
(but no such surfaces exist in general). There is a spatial metric or squared line element
\be
h_{\mu\nu}dx^\mu dx^\nu\equiv e_\mu^a e_\nu^a dx^\mu dx^\nu,
\ee
where $h_{\mu\nu}=e_\mu^a e_\nu^a\equiv e_\mu^a e_\nu^b \eta_{ab}$ and $\eta_{ab}$ is the standard internal
spatial  metric,
given by the identity matrix or $\eta_{ab}=\delta_{ab}$. (Note the use of notation like
$v^aw^a=v^aw^b\eta_{ab}$ with summation convention, as a way of contracting internal spacelike indices, and
a similar convention for the case of lower indices.) ``Inverse'' spatial metrics $h^{\mu\nu}$ and
$\eta^{ab}$ with upper indices can be defined likewise, but notice that the ambient spacetime metrics are
degenerate and not truly inverses of each other; instead
$h_{\mu\nu}h^{\nu\lambda}=\delta_\mu^\lambda-e_\mu^0e_0^\lambda$.  We assume that both the ambient spatial
metrics are positive semidefinite. It may be tempting to combine these timelike and spacelike partial metric
tensors into a single spacetime metric, but because of the lack of Lorentz invariance, this is not
necessary, nor would it be uniquely defined\cite{Kunzle1972,Son2013} (line elements of time and space have
different dimensions;
there is no universal scale of speed). Therefore such a metric will {\em not} be used, and  in general we
do not raise or lower any indices (occasionally we do so for internal spacelike indices using $\eta_{ab}$
or $\eta^{ab}$).

One could make different choices of the one-forms $e_\mu^\alpha$ that differ by a linear transformation of
the internal indices $\alpha$. Because $\alpha=0$ has been singled out, and because of the choice of
internal metric on the space components which we may as well fix, the only possible transformations are
SO($d$) rotations on the internal space-like indices $a$, $b$ only (we neglect improper rotations of
negative determinant). These rotations act as internal gauge transformations, as in a Yang-Mills gauge
theory. To make expressions containing partial derivatives covariant under such transformations, we need
a gauge field or ``spin connection'' $\omega_{\mu\hphantom{\alpha}\beta}^{\hphantom{\mu}\alpha}$ (an
example of its use will appear in a moment). In the present case, only the internal
space-like components $\omega_{\mu\hphantom{a}b}^{\hphantom{\mu}a}$ are nonzero. In view of the standard
Euclidean metric on internal spacelike indices, we can raise or lower an index $a$ or $b$, and it makes
sense to say that the spin connection is antisymmetric on its internal indices [it is in the Lie algebra
of SO($d$)]. For $d=2$, the spin connection is effectively a pseudoscalar on the internal indices.

We will also need a Christoffel connection with components (or Christoffel symbols)
$\Gamma_{\hphantom{\mu}\nu\lambda}^{\mu}$ in order to write
covariant derivatives in spacetime. As an example of the use of the two connections, the covariant
derivative of the one-forms is
\be
\nabla_\mu e_\nu^\alpha=\partial_\mu
e_\nu^\alpha+\omega_{\mu\hphantom{\alpha}\beta}^{\hphantom{\mu}\alpha}e_\nu^\beta-
\Gamma_{\hphantom{\lambda}\mu\nu}^{\lambda}e_\lambda^\alpha.
\ee
We do {\em not} impose the symmetry
condition $\Gamma_{\hphantom{\mu}\nu\lambda}^{\mu}=\Gamma_{\hphantom{\mu}\lambda\nu}^{\mu}$, which
means that our spacetime generally has torsion; the torsion tensor is
$T^\mu_{\nu\lambda}=\Gamma_{\hphantom{\mu}\nu\lambda}^{\mu} -
\Gamma_{\hphantom{\mu}\lambda\nu}^{\mu}$. It frequently appears with an upper internal index $\alpha$ in
place of $\mu$. We call the $\alpha=0$ components of $T^\alpha_{\mu\nu}$ the timelike torsion, and the
$\alpha=a$ components the spacelike torsion.

The one-forms, spin connection, and Christoffel connection are not necessarily independent. We impose the
requirement that
\be
\nabla_\mu e_\nu^\alpha=0,
\label{covconsteq}
\ee
so the vielbeins (and their inverses) are covariantly constant. When we come to varying an action, we must
specify which variables are viewed as independent, and in the first part of the paper, we choose
to view the vielbein and spin connection as the independent variables that describe the spacetime geometry.
The covariant constancy equation can be solved for the Christoffel symbols, and the torsion is
\begin{equation}
T_{\mu\nu}^\ssc=\partial_\mu e_\nu^\ssc-\partial _\nu
e_\mu^\ssc+\omega_{\mu\hphantom{\alpha}\beta}^{\hphantom{\mu}\alpha}
e_\nu^\beta-\omega_{\nu\hphantom{\alpha}\beta}^{\hphantom{\nu}\alpha} e_\mu^\beta. \label{cartan}
\end{equation}
The components of the timelike torsion are essentially the curl (or exterior
derivative) of the one-form $e_\mu^0$; the vanishing of these is precisely the condition above
for the existence of an absolute time coordinate. Later in the paper, we will also make use of a different
point of view, in which the vielbeins $e_\mu^\alpha$ and what we will call the reduced
torsion $\Tt_{\mu\nu}^a$ (a part of the spacelike torsion that is independent of the vielbeins;
the timelike torsion is fully determined by the timelike vielbeins in any case) will be
viewed as the independent variables; using the covariant-constancy equation, one can express the
Christoffel symbols and the spin connection in terms of these. As these expressions are more
lengthy, they are given in Appendix \ref{appB}.

Our actions will involve integration over spacetime, and we will need to use a volume-form or
measure for the integration. This is simply constructed from the time-like and space-like metrics above,
and will be written as $d^{d+1}x\,\widehat{\sqrt{g}}$ as usual, where $\widehat{\sqrt{g}}$ (which does
not transform as a scalar) is defined (for $d=2$, but other dimensions are similar) by
\begin{equation}
\widehat{\sqrt{g}}=\frac{1}{6}\widehat{\epsilon}^{\mu\nu\lambda}\epsilon_{\ssc\ssh\sss}e_\mu^\ssc
e_\nu^\ssh e_\lambda^\sss,
\end{equation}
which clearly is simply the determinant of the matrix with entries $e_\mu^\alpha$;
the non-degeneracy condition implies that it is non-zero at all spacetime points, and we assume it
is positive. The ambient spacetime epsilon symbol (not tensor) $\widehat{\epsilon}^{\mu\nu\lambda}$
is defined in any
coordinate system (again for $d=2$) by $\widehat{\epsilon}^{012}=1$, and the internal one (which is an
invariant tensor for the internal transformations, essentially spatial rotations, that we use)
${\epsilon}^{\alpha\beta\gamma}$ likewise. The
lower-index ones $\epsilon_{\alpha\beta\gamma}$ and $\check{\epsilon}_{\mu\nu\lambda}$ are defined in
the same way. The notation with a hat used here will indicate throughout that the object on which it
appears is a tensor density, rather than a tensor, which transforms under coordinate transformation
with an additional determinantal factor (as $\widehat{\sqrt{g}}$ does) compared with a tensor with the
same ambient spacetime indices; alternatively, a tensor density divided by $\widehat{\sqrt{g}}$ transforms
as a tensor. (The lower ambient index epsilon symbol with the check symbol transforms inversely to the
upper index one.) We note the useful relation
\begin{equation}
\Gamma^{\nu}_{\hphantom{\nu}\mu\nu}=\frac{1}{\widehat{\sqrt{g}}}\partial_\mu\widehat{\sqrt{g}}.
\end{equation}
Without the hat, $\epsilon^{\mu\nu\lambda}=\epsilon^{\alpha\beta\gamma}e_\alpha^\mu e_\beta^\nu
e_\gamma^\lambda$ is a tensor, and similarly for $\epsilon_{\mu\nu\lambda}$, which is the volume three-form
written in components.
We also sometimes use the two-index epsilon symbol $\epsilon^{ab}=\epsilon^{0ab}$, and likewise for
lower indices, which is natural in view of the singling out of timelike components, and can even be done
for the ambient versions as $\epsilon^{\mu\nu}=e^0_\lambda\epsilon^{\lambda\mu\nu}$.

We also define here the Riemann curvature tensor, although it will not appear much in this paper.
This can be obtained \cite{Carroll2004} from the commutator of two covariant derivatives
$[\nabla_\mu,\nabla_\nu]$ applied to a vector field with an
internal index, say $v^\alpha$. (By covariant constancy of the vielbein which can be used to convert
indices, this determines the Riemann tensor in general.) We have
\be
R_{\mu\nu\hphantom{\alpha}\beta}^{\hphantom{\mu\nu}\alpha} = \partial_\mu
\omega_{\nu\hphantom{\alpha}\beta}^{\hphantom{\nu}\alpha}
-\partial_\nu \omega_{\mu\hphantom{\alpha}\beta}^{\hphantom{\mu}\alpha}
+\omega_{\mu\hphantom{\alpha}\gamma}^{\hphantom{\mu}\alpha}
\omega_{\nu\hphantom{\gamma}\beta}^{\hphantom{\mu}\gamma}
-\omega_{\nu\hphantom{\alpha}\gamma}^{\hphantom{\mu}\alpha}
\omega_{\mu\hphantom{\gamma}\beta}^{\hphantom{\mu}\gamma},
\ee
and so vanishes unless $\alpha=a$, $\beta=b$. In the case $d=2$ of interest in this paper, the
non-vanishing components reduce to
\be
R_{\mu\nu\hphantom{a}b}^{\hphantom{\mu\nu}a} = \partial_\mu \omega_{\nu\hphantom{a}b}^{\hphantom{\nu}a}
-\partial_\nu \omega_{\mu\hphantom{a}b}^{\hphantom{\mu}a},
\ee
which is effectively the curl of the single one-form $\omega_{\mu\hphantom{1}2}^{\hphantom{\mu}1}$,
similar to the case in Ref.\ \onlinecite{Wen1992}.

Finally, we note that the variation of spacetime tensors under diffeomorphisms is given by the Lie
derivative $\mathcal{L}$. Under a diffeomorphism generated by the vector field $\xi$, we have for scalar
functions
\begin{equation}
\mathcal{L}_\xi f=\xi^\nu\partial_\nu f,
\end{equation}
for vectors
\begin{equation}
\mathcal{L}_\xi V^\mu=\xi^\nu\partial_\nu V^\mu-V^\nu\partial_\nu\xi^\mu,
\end{equation}
and for one-forms
\begin{equation}
\mathcal{L}_\xi W_\mu=\xi^\nu\partial_\nu W_\mu+W_\nu\partial_\mu\xi^\nu.
\end{equation}
Although we will not need it here, the generalization to higher rank tensors is obtained by demanding
that $\mathcal{L}$ satisfies the Liebnitz rule.

In addition to these geometric structures in spacetime, we will also use a U($1$) gauge potential $A_\mu$,
with field strength $F_{\mu\nu}=\partial_\mu A_\nu-\partial_\nu A_\mu$ as usual. For covariant derivatives
of fields that carry U($1$) charge, as in the following section, $\nabla_\mu$ will denote the
fully-covariant derivative that includes the vector potential.

\subsection{Action for matter}
\label{sec:matter}

We now consider actions for non-relativistic matter fields, as an illustration of the use of the above
background fields, and to check that the variations with these background fields produce the correct
conserved currents. In relation to this, the vielbeins play the role of gauge potentials that enable us,
in some sense, to gauge translation invariance, and so variations with respect to them produce the
corresponding covariantly conserved currents and densities, just as varying the electromagnetic gauge
potential produces the conserved electric current/density (that is, satisfying the continuity equation).
As an example, let us consider a minimally-coupled second-quantized action for spinless bosons or fermions
in flat spacetime,
\begin{align}
S&=\int d^{d+1}x \big(i\varphi^\dag D_0\varphi -\frac{1}{2m}(D_i\varphi)^\dag
D_i\varphi\big)
\nonumber \\
&-\frac{1}{2}\int d^{d+1}x\,d^{d+1}y\, V(x-y)\varphi^\dag(x)\varphi^\dag(y)\varphi(x)\varphi(y),
\end{align}
where $\varphi$ is a scalar field (either commuting or anticommuting, for bosons or fermions respectively),
$D_\mu=\partial_\mu-iA_\mu$ is the gauge-covariant derivative, and
$V$ is an interaction potential in spacetime. For a general spacetime we obtain the covariant action
\begin{align}
S=&\int d^{d+1}x\widehat{\sqrt{g}}\Big(\frac{1}{2}ie^\mu_0(\varphi^\dag
\overleftrightarrow{\nabla}_\mu\varphi)-\frac{1}{2m}e^\mu_ae^\nu_a(\nabla_\mu\varphi)^\dag
\nabla_\nu\varphi\Big.\nonumber\\
&+\frac{1}{2}\int d^{d+1}y \widehat{\sqrt{g}}\,V(x, y)
\varphi^\dag(x)\varphi(x)\varphi^\dag(y)\varphi(y)\Big).\label{microaction}
\end{align}
The expression for the interaction term containing $V(x,y)$ requires some care. For the case of contact
interactions, where
the interaction potential $V$ is given by a differential operator acting on a delta function, there are
no issues as the delta function is already a scalar density; we need simply to replace the derivatives
acting upon it with covariant derivatives, contracted using the spacelike metric. We note that such contact
interactions may be taken to be independent of the spacelike torsion $T^a_{\mu\nu}$.

For finite-range instantaneous interactions, we need to generalize the notion of spacelike distance to
curved spacetime with torsion. There are two conceptual difficulties here. First, if the time-like torsion
$T^0_{\mu\nu}$ is not identically zero in a region of spacetime, there does not exist an absolute time
variable defined in that region as we mentioned before, and so hypersurfaces of constant absolute time do
not exist. However, that condition, which says that $e^0_\mu$ is an exact differential, is more restrictive
than is necessary for this purpose, and in general, according to a theorem of
Frobenius\cite{Carroll2004,Lang2002},
hypersurfaces whose tangent vectors $e^\mu_a$ are orthogonal to $e_\mu^0$ at each point exist if and only
if the weaker condition
\begin{equation}
e^\mu_ae^\nu_bT^0_{\mu\nu}=0,
\end{equation}
holds throughout a region, that is when the tangent vector fields $e^\mu_a$ are integrable. For $d=2$, this
expression can also be written as
\be
\epsilon^{\mu\nu\lambda}e^0_\lambda T^0_{\mu\nu}=0.
\label{hypcond}
\ee
For general $d$, one can write this simply as $e^0_{[\lambda}T^0_{\mu\nu]}=0$, where the square brackets
surrounding indices means antisymmetrization.  In this form, for all $d$, the (dual version of the)
Frobenius theorem says equivalently that the condition is satisfied if and only if $e_\mu^0$
obeys an equation of the form $e_\mu^0 dx^\mu =\psi dw$ for some scalar functions $\psi(x)$,
$w(x)$; then the spacelike hypersurfaces are surfaces of constant $w$.

Second, in the presence of
torsion we must distinguish between spacelike \emph{geodesics} --- paths $r^\mu(\lambda)$ that satisfy
the both geodesic equation\cite{Carroll2004}
\begin{align}
0&=\frac{d^2r^\mu}{d\lambda^2}+\Gamma^\mu_{\hphantom{\mu}\nu\rho}\frac{dr^\nu}{d\lambda}
\frac{dr^\rho}{d\lambda}
\end{align}
and the spacelike constraint
\begin{align}
0&=e_\mu^0\frac{dr^\mu}{d\lambda}
\end{align}
---and spacelike paths of minimal distance. The geodesics are those paths which parallel transport their
tangent vectors, and in the absence of torsion, these coincide with paths of minimal distance. This can be
seen by examining Eq. (\ref{christoffels}) for the Christoffel symbols, and noting that the Euler-Lagrange
equation for minimization of spacelike distance depends only on the contribution of the spacelike metric to
the connection. Here we will work with spacelike geodesics because they are easier to construct.

Given a point $x^\mu$ on our manifold, we denote by $r^\mu_x(v^a,\lambda)$ the parametrized geodesic
satisfying the initial condition that its tangent is along a spacelike vector $v^a$
\begin{align}
r_x^\mu(v^a,0)&=x^\mu, \\
\left.\frac{dr_x^\mu}{d\lambda}\right|_{\lambda=0}&=v^ae^\mu_a
\end{align}
Because of the possible reparametrizations of $\lambda$, $v^a$ is only defined up to a scalar factor.
If our manifold is sufficiently well-behaved (i.e.\ geodesically complete), we may take
$\lambda\in\left(-\infty,\infty\right)$. Since we do not have a notion of spacelike hypersurfaces, we must
make do with the set of all points connected to $x^\mu$ by spacelike geodesics. More formally, we consider
the open sets $U_x$ defined by
\begin{equation}
U_x=\left\{r_x^\mu(v^a,\lambda)\right\}.
\end{equation}
Note that the sets $U_x$ are the images of the exponential map acting on the set of spacelike tangent
vectors at $x$, and hence they are proper $d$-dimensional submanifolds of spacetime\cite{Helgason2001}.
For each $y\in U_x$, we may then define the distance
\begin{equation}
d_x(y)=\left|\int_0^{\lambda_0}{d\lambda\sqrt{h_{\mu\nu}\frac{dr_x^\mu}{d\lambda}\frac{dr_x^\nu}{d\lambda}}}
\right|.
\end{equation}
Note that $d_x(y)=d_y(x)$ by the uniqueness of solutions to the geodesic equation. Using this distance,
we may form the covariant interaction term
\begin{align}
S_{int}=\frac{1}{2}\int d^{d+1}x\,\widehat{\sqrt{g}}\int{d^{d+1}y\,\widehat{\sqrt{g}}}&
\left[\chi_{U_{x}}(y)V\left(d_x(y)\right)\right.\nonumber \\
&\left.\times\varphi^\dag(x)\varphi(x)\varphi^\dag(y)\varphi(y)\right],
\end{align}
where $\chi_{U_x}$ is the characteristic function of the set $U_x$. This expression is rather cumbersome,
and we will not make explicit use of it in the remainder of this work. However, we note that in the absence
of torsion, it reduces to a straightforward generalization of the interaction term constructed in Ref.\
\onlinecite{Bradlyn2012}.

Returning to our expression Eq.\ (\ref{microaction}) for the microscopic action, it is illuminating to
assign independent meaning to certain components of $e_\mu^\ssc$, or more precisely,
to $\delta e_\mu^\ssc=e_\mu^\ssc-\delta_\mu^\ssc$. By examining the action, we see that $\delta e_0^0$
enters exactly as the artificial gravitational potential $\psi$ introduced by Luttinger for calculating
thermal response functions \cite{Luttinger1964,Cooper1997}; when this is the only nonzero component of
$e$, it multiplies the energy density. This is consistent with the standard Newtonian approximation to
gravity in the general relativity literature\cite{Carroll2004}. Notice that the spatial components $e_i^a$
enters similarly as the matrix $\Lambda$ of Ref.\
\onlinecite{Bradlyn2012}. This is no accident: the matrix $\Lambda$ presented there is very much just
these components of the vielbein, and to first order $\delta e_i^a$ are the matrices $\lambda_i^a$. Because
it couples longitudinally to the heat current, $\delta e^0_i$ can be interpreted as the ``gravitomagnetic
vector potential'' mentioned recently in the literature\cite{Nomura2012,Ryu2012}.

Next, we will outline the general procedure for obtaining equations of motion and the various currents from
an action, and obtain the conservation laws for the currents from the invariance properties. We use
the action above (or the version with $V=0$) as an example with which to check the results. Given an
action $S$ involving the background fields and a scalar field $\varphi$, and now taking the vielbeins and
spin connection as the independent background fields, we can consider the variations
\begin{align}
\varphi&\rightarrow\varphi+\delta\varphi ,\\
A_\mu&\rightarrow A_\mu+\delta A_\mu, \\
e_\mu^\alpha&\rightarrow e_\mu^\alpha+\delta e_\mu^\alpha, \\
\omega_{\mu\hphantom{a}b}^{\hphantom{\mu}a}&\rightarrow\omega_{\mu\hphantom{a}b}^{\hphantom{\mu}a}
+\delta\omega_{\mu\hphantom{a}b}^{\hphantom{\mu}a},
\end{align}
to obtain
\begin{eqnarray}
\delta S&=&\int d^{d+1} x\, \left[\frac{\delta S}{\delta e^{\mu}_\ssc}\delta e^{\mu}_\ssc
+\frac{\delta S}{\delta \varphi}\delta \varphi\right.\non\\
&&{}\left.+\frac{\delta S}{\delta A_\mu}\delta A_\mu
+\frac{\delta S}{\delta
\omega_{\mu\hphantom{a}b}^{\hphantom{\mu}a}}\delta\omega_{\mu\hphantom{a}b}^{\hphantom{\mu}a}\right].
\end{eqnarray}
For the equations of motion of the matter field, the variation of the action with the background fields
fixed is set to zero, and so the equations of motion are
\begin{equation}
\frac{\delta S}{\delta\varphi}=0. \label{eom}
\end{equation}

Now we define several currents. These are the number current (with components $\mu=0$ for density and
$\mu=i$ for spatial current)
\begin{equation}
J^\mu=\frac{1}{\widehat{\sqrt{g}}}\frac{\delta S}{\delta A_\mu},
\end{equation}
and analogously the energy-momentum-stress current (or tensor)
\be
\tau^\mu_{\hphantom{\mu}\alpha}=-\frac{1}{\widehat{\sqrt{g}}}\frac{\delta S}{\delta e_\mu^\alpha}.\ee
The latter contains the energy current $J^\mu_E=\tau^\mu_0$ as the $\alpha=0$ components,
and the momentum current as the $\alpha=a$ components, of which the $\mu=0$ component is momentum density,
and the $\mu=i$ components are the momentum flux or (essentially) the stress. Finally, there is the spin
current
\begin{equation}
J^{\hphantom{S}\mu\hphantom{a}b}_{S\hphantom{\mu}a}=\frac{1}{\widehat{\sqrt{g}}}\frac{\delta S}{\delta
\omega_{\mu\hphantom{a}b}^{\hphantom{\mu}a}}. \label{spincurrentdef}
\end{equation}
In $d=2$ dimensions, the spin current is antisymmetric in $a$, $b$, and those indices can be dropped.

Next we obtain the conservation laws for these currents from the local symmetries.
Considering first an infinitesimal $U(1)$ gauge transformation
\begin{align}
\delta\varphi&=i\varphi\theta, \\
\delta A_\mu&=\partial_\mu\theta, \\
\delta e_\mu^\alpha&=0, \\
\delta\omega_{\mu\hphantom{a}b}^{\hphantom{\mu}a}&=0,
\end{align}
with a scalar function $\theta(x)$,
we find, after using the equations of motion Eq. (\ref{eom}) and the fact that the variation of the action
under a symmetry transformation is by definition zero, the number current conservation law
\begin{equation}
\frac{1}{\widehat{\sqrt{g}}}\partial_\mu (\widehat{\sqrt{g}} J^\mu)=\nabla_\mu
J^\mu-T^{\nu}_{\nu\mu}J^\mu=0.
\end{equation}

Next we wish to examine local space and time translations. We can do this in one fell swoop by looking at
how the action changes under arbitrary infinitesimal diffeomorphisms $x^\mu\rightarrow x^\mu+\xi^\mu$,
where $\xi^\mu(x)$ is a vector field. This has the
effect of modifying all fields by their Lie derivatives as pointed out above. However, because the
Lie derivative is not explicitly covariant, it is useful to modify it to also include a well-chosen
$U(1)$ gauge transformation and an internal rotation. That is, to the Lie derivative of charged fields we
add an additional gauge transformation by the amount $\xi^\mu A_\mu$, and to the Lie derivative of fields
with an internal index we add an additional internal rotation by the amount
$\xi^\mu\omega_{\mu\hphantom{a}b}^{\hphantom{\mu}a}$. We are free to do this since these transformations
are themselves symmetries of the action. A short calculation shows that the field variations are then
given by the covariant Lie derivatives
\begin{align}
\delta\varphi&=\xi^\mu\nabla_\mu\varphi, \\
\delta A_\mu&=\xi^\nu F_{\nu\mu}, \\
\delta e_\mu^\alpha&=e_\nu^\alpha\nabla_\mu\xi^\nu-T^\alpha_{\mu\nu}\xi^\nu, \\
\delta\omega_{\mu\hphantom{a}b}^{\hphantom{\mu}a}&=\xi^\nu R_{\mu\nu\hphantom{a}b}^{\hphantom{\mu\nu}a},
\end{align}
yielding, after an application of the equations of motion, the energy-momentum conservation law
\begin{align}
\nabla_\mu\tau^\mu_{\hphantom{\mu}\alpha}-T^\lambda_{\lambda\mu}\tau^\mu_{\hphantom{\mu}\alpha}
=-e^\nu_\alpha\left(J^\mu
F_{\mu\nu}+J_{S\hphantom{\mu}a}^{\hphantom{S}\mu\hphantom{a}b}R_{\mu\nu\hphantom{a}b}^{\hphantom{\mu\nu}a}
+\tau^\mu_{\hphantom{\mu}\beta}T^\beta_{\mu\nu}\right).\label{stresscont}
\end{align}
The contribution on the right-hand side of the form spin current times Riemann curvature is a known effect
that corresponds to a force on spinning bodies due to curvature.

Finally, using an infinitesimal internal rotation
\begin{align}
\delta e_\mu^a&=\Omega^a_{\hphantom{a}b}e_\mu^b, \\
\delta\omega_{\mu\hphantom{a}b}^{\hphantom{\mu}a}&=\Omega^a_{\hphantom{a}c}
\omega_{\mu\hphantom{a}b}^{\hphantom{\mu}c}-\Omega^c_{\hphantom{c}b}
\omega_{\mu\hphantom{a}c}^{\hphantom{\mu}a}
-\partial_\mu\Omega^a_{\hphantom{a}b},
\end{align}
with $\Omega^a_{\hphantom{a}b}(x)$ an arbitrary antisymmetric matrix function, we find that the
antisymmetric part of the stress tensor satisfies
\begin{equation}
\epsilon^a_{\hphantom{a}b}\tau^b_{\hphantom{b}a}=\epsilon^a_{\hphantom{a}b}\left(\nabla_\mu
J_{S\hphantom{\mu}a}^{\hphantom{S}\mu\hphantom{a}b}-T^\nu_{\nu\mu}
J_{S\hphantom{\mu}a}^{\hphantom{S}\mu\hphantom{a}b}\right), \label{antisym}
\end{equation}
where
\begin{equation}
\tau^b_{\hphantom{b}a}=e^b_\mu\tau^\mu_{\hphantom{\mu}a}.
\end{equation}
This can also be viewed as the conservation law for the spin current.

With this formalism established, we can now proceed to identify these conserved currents with physical
quantities. Here, we focus on the flat space $e=Id,\omega=0$ expressions of the currents, deferring
discussion of the more general case (and the associated ``contact'' terms) until Sec. \ref{respsec}.
We will also set the interaction potential $V$ to zero for brevity. Using the action
Eq.\ (\ref{microaction}), we find for the number current
\begin{align}
J^\mu&=\left.\frac{1}{\widehat{\sqrt{g}}}\frac{\delta S}{\delta A_\mu}\right|_{e=Id,\omega=0} \nonumber \\
&=\delta^{\mu}_0\varphi^\dag\varphi-\frac{i}{2m}\left(\varphi^\dag
D_i\varphi-(D_i\varphi)^\dag\varphi\right)\delta^\mu_i,
\end{align}
as expected for a charged field. For the energy-momentum-stress tensor, things are quite a bit more
complicated, but we eventually find
\begin{align}
\tau^\mu_{\hphantom{\mu}\alpha}=&-\left.\frac{1}{\widehat{\sqrt{g}}}\frac{\delta S}{\delta
e_\mu^\alpha}\right|_{e=Id,\omega=0} \nonumber \\
=&\frac{i}{2}\left(\delta^\mu_0\delta^\nu_\alpha-\delta^\mu_\alpha \delta^\nu_0\right)\left(\varphi^\dag
D_\nu\varphi-(D_\nu\varphi)^\dag\varphi\right) \nonumber \\
&-\frac{1}{2m}\left(\delta^\lambda_\alpha\delta^\nu_a\delta^\mu_a
+\delta^\nu_\alpha\delta^\lambda_a\delta^\mu_a
-\delta^\mu_\alpha\delta^\nu_a\delta^\lambda_a\right)(D_\lambda\varphi)^\dag D_\nu\varphi.
\end{align}
Unpacking terms, we see (after using the equations of motion to eliminate time derivatives) that the
$\alpha=0$ components of $\tau$ give the energy density and spatial energy current consistent with Ref.\
\onlinecite{Cooper1997}, while the $\alpha=a$ components give \emph{minus} the momentum density and
stress tensor consistent with Ref.\ \onlinecite{Bradlyn2012}, plus an additional term
$\frac{1}{4m}D^2J^0\delta^{\mu}_{\alpha}$ due to operator ordering (c.f.\ Ref.\ \onlinecite{Gromov20141}).
Finally, the spin current is zero because the action does not contain the spin connection.

Readers will have noticed that there is no chemical potential in our action. That is because we work in
the canonical ensemble with a fixed particle number $N$. $N$ is the flux of $J^\mu$ across an arbitrary (in
principle spacelike) section $\cal A$; as $J^\mu$ obeys a covariant continuity equation, $N$ is invariant
under small changes in the section. Precisely, the flux can be written (for $d=2$; other dimensions are
similar)
\be
N=\int_{\cal A} \epsilon_{\mu\nu\lambda}J^\mu dx^\nu dx^\lambda,
\ee
and we note that $\epsilon_{\mu\nu\lambda}J^\mu$ is the set of components of a two-form, and the two-form
(in general, a $d$-form) can be integrated over a $d$-surface without any use of the metric. $N$ is
invariant under small changes in the section because conservation implies
$\partial_{[\rho}\epsilon_{\nu\lambda]\mu}J^\mu=0$. Thus
classically, the expression for $N$ has to be imposed as a constraint; quantum mechanically, in an operator
formalism, one uses
only states that obey this as an initial condition, that is preserved by time evolution; it can be imposed
in a functional integral treatment by introducing an integration over an additional variable (actually a
gauge potential) to make a functional $\delta$-function. In general, the effect of the global constraint
is only felt globally, and if we eventually consider response functions in flat spacetime with a
translation invariant system, the effect only shows up at zero wavevector $\bf k$ in responses that couple
to the particle number. For quantities of interest we can take the limit as ${\bf k}\to0$ instead of ${\bf
k}=0$ when it makes a difference. Thus in practice, when studying local behavior in a large system, we
will simply ignore the number constraint. (If desired it can be incorporated along the lines mentioned.)

\section{Induced action: generalities}
\label{sec:gen}

Our goal is to find the most general induced bulk action
\begin{equation}
S^{\mathrm{eff}}[A_\mu,e_\mu^\alpha,\omega_{\mu\hphantom{a}b}^{\hphantom{\mu}a}]\nonumber
\end{equation}
one could obtain for a system that is gapped in the bulk (i.e.\ a topological phase ) after integrating
out the matter fields. As indicated, this induced action is a functional of the
electromagnetic potential
$A_\mu$, the vielbeins $e_\mu^\alpha$, and the spin connection
$\omega_{\mu\hphantom{a}b}^{\hphantom{\mu}a}$, which for now we continue to use as the independent
background fields. We can expand this functional as the integral of a sum of local terms,
\begin{align}
S^{\mathrm{eff}}=&\int d^{d+1}x\,\widehat{\sqrt{g}}\left[\mathcal{L}^{(0)}(A_\mu,e_\mu^\alpha,
\omega_{\mu\hphantom{a}b}^{\hphantom{\mu}a})\right.\nonumber\\
&\left.+\mathcal{L}^{(1)}(A_\mu,e_\mu^\alpha,\omega_{\mu\hphantom{a}b}^{\hphantom{\mu}a})
+\dots\right].
\end{align}
Each such $\mathcal{L}^{(n)}$ is a function of the external fields and their derivatives, and each {\em
integral} must be invariant under coordinate transformations, internal rotations, and electromagnetic
gauge transformations, up to boundary terms. Very generally, these terms can be divided into two
categories. The first category, which we term ``locally invariant,'' consists of those terms in which the
$\mathcal{L}^{(n)}$ themselves are invariant under all the aforementioned symmetry transformations.
These terms can be written as polynomials in strictly covariant quantities such as the vielbein
$e_\mu^\alpha$, the electromagnetic field strength $F_{\mu\nu}$, the torsion $T^\alpha_{\mu\nu}$, the
curvature $R_{\mu\nu\hphantom{\alpha}\beta}^{\hphantom{\mu\nu}\alpha}$, and their (covariant)
derivatives, with appropriate index contractions. All such terms can be combined into one action
\begin{equation}
S^{\mathrm{loc}}[e_\mu^\alpha,F_{\mu\nu},T^0_{\mu\nu},T^a_{\mu\nu},
R_{\mu\nu\hphantom{\alpha}\beta}^{\hphantom{\mu\nu}\alpha}] \label{locfunc}
\end{equation}
which is a functional of these covariant tensors and their covariant derivatives.
The second category consists of the remaining integrands $\mathcal{L}^{(n)}$ which are not
invariant under all the transformations, but invariant only up to total derivatives for at least one
type of transformation.
The integrals of such terms are invariant only up to boundary contributions, and in general
invariance of the total action necessitates the existence of gapless edge degrees of freedom.
An example of such a term in $2+1$ dimensions is the familiar U($1$) Chern-Simons term
\begin{equation}
\mathcal{L}^{(\mathrm{CS})}=\epsilon^{\mu\nu\lambda}A_\mu\partial_\nu A_\lambda
\end{equation}
which is not manifestly invariant under a gauge transformation, but changes by a total derivative.
A locally-covariant term can be multiplied by an arbitrary function of position, and would still be
covariant, and this could occur due to changes of parameters with position in the microscopic action.
For a Chern-Simons-type term, on the other hand, this cannot be done as it spoils the invariance up to
a total derivative. This implies that the coefficient in a Chern-Simons term should usually be robust
against changes in the parameters in the underlying ``microscopic'' action throughout a topological
phase that respects the symmetries in question; if it were not, varying microscopic
parameters in spacetime would lead to changes in the coefficient and so violate invariance of the induced
action. (This field-theoretic argument, often formulated as the non-renormalization of the coefficients
in Chern-Simons-type terms, deserves to be more familiar in condensed matter physics.)
Further Chern-Simons-type terms that can occur in our theories in $2+1$ dimensions are
the first and second Wen-Zee terms\cite{Wen1992}
\begin{align}
\mathcal{L}^{(\mathrm{WZ}1)}&=\epsilon^{\mu\nu\lambda}\omega_\mu\partial_\nu A_\lambda, \\
\mathcal{L}^{(\mathrm{WZ}2)}&=\epsilon^{\mu\nu\lambda}\omega_\mu\partial_\nu\omega_\lambda,
\end{align}
as well as the gravitational Chern-Simons term
\begin{equation}
\mathcal{L}^{(\mathrm{GCS})}=\epsilon^{\mu\nu\lambda}\left(\Gamma^\rho_{\mu\sigma}
\partial_{\nu}\Gamma^{\sigma}_{\nu\rho}+\frac{2}{3}\Gamma^{\rho}_{\mu\sigma}
\Gamma^{\sigma}_{\nu\theta}\Gamma^{\theta}_{\lambda\rho}\right),
\end{equation}
(here we use $\theta$ as an ambient index---as usual with this sort of thing, one very quickly runs out
of Greek letters). These terms must be treated individually on a
case-by-case basis.

Given such an induced action, one can effect the functional derivatives as in the previous section
to compute the expectation values of currents in the presence of a given background configuration of the
fields $A_\mu$, $e_\mu^\alpha$, and $\omega_{\mu\hphantom{a}b}^{\hphantom{\mu}a}$. In particular,
we can define the number current
\begin{equation}
J^\mu=\frac{1}{\widehat{\sqrt{g}}}\frac{\delta S^{\mathrm{eff}}}{\delta A_\mu}, \label{avJ}
\end{equation}
the energy-momentum-stress tensor
\begin{equation}
\tau^\mu_{\hphantom{\mu}\alpha}=-\frac{1}{\widehat{\sqrt{g}}}\frac{\delta S^{\mathrm{eff}}}{\delta
e_\mu^\alpha},\label{avtau}
\end{equation}
and the spin current
\begin{equation}
J^{\hphantom{S}\mu\hphantom{a}b}_{S\hphantom{\mu}a}=\frac{1}{\widehat{\sqrt{g}}}\frac{\delta
S^{\mathrm{eff}}}{\delta
\omega_{\mu\hphantom{a}b}^{\hphantom{\mu}a}}, \label{avJs}
\end{equation}
where we use the same notation for the currents and their expectation values, as we expect the meaning
to be clear from context. The conservation laws obeyed by the currents are the same as in the previous
section.

In computing the currents from the induced action, we see that the
contributions from the locally invariant terms and from the Chern-Simons-type terms have very
different structure.
The contributions of the locally-invariant terms to the currents have the forms
\begin{align}
J^\mu_{\mathrm{loc}}=&\frac{1}{\widehat{\sqrt{g}}}\partial_\lambda\left(
\frac{\delta S^{\mathrm{loc}}}{\delta F_{\mu\lambda}}\right), \\
\tau^{\mu}_{\hphantom{\mu}\alpha,\mathrm{loc}}=&-\frac{1}{\widehat{\sqrt{g}}}
\frac{\delta S^{\mathrm{loc}}}{\delta e_\mu^a}-\frac{1}{\widehat{\sqrt{g}}}\partial_\lambda
\left(\frac{\delta S^{\mathrm{loc}}}{\delta
T^\alpha_{\mu\lambda}}\right)\nonumber\\
&{}-\frac{1}{\widehat{\sqrt{g}}}\omega_{\lambda\hphantom{\alpha}\beta}
^{\hphantom{\nu}\beta}\frac{\delta S^{\mathrm{loc}}}{\delta T^\beta_{\mu\lambda}}, \\
J^{\hphantom{S}\mu\hphantom{a}b}_{S\hphantom{\mu}a\hphantom{b},\mathrm{loc}}=&\frac{1}{\widehat{\sqrt{g}}}
e_\lambda^b\frac{\delta S^{\mathrm{loc}}}{\delta
T_{\mu\lambda}^a}+\frac{1}{\widehat{\sqrt{g}}}\partial_{\lambda}
\left(\frac{\delta S^{\mathrm{loc}}}{\delta
R_{\mu\lambda\hphantom{a}b}^{\hphantom{\mu\lambda}a}}\right)\nonumber\\
&{}+\frac{1}{\widehat{\sqrt{g}}}\omega_{\lambda\hphantom{a}c}^{\hphantom{\nu}a}
\frac{\delta S^{\mathrm{loc}}}{\delta
R_{\mu\lambda\hphantom{a}b}^{\hphantom{\mu\lambda}c}}-\frac{1}{\widehat{\sqrt{g}}}
\omega_{\lambda\hphantom{a}b}^{\hphantom{\nu}c}
\frac{\delta S^{\mathrm{loc}}}{\delta R_{\mu\lambda\hphantom{a}c}^{\hphantom{\mu\lambda}a}}.
\end{align}
Each functional derivative here is taken with the remaining arguments in Eq. (\ref{locfunc}) held fixed.
There are two types of terms that appear here. The first type occurs as the first term on the right-hand
side of the last two equations, which enter because $e_\mu^\alpha$ and
$\omega^{\hphantom{\mu}a}_{\mu\hphantom{a}b}$ can appear in $S^{\mathrm{loc}}$ without derivatives.
The remaining terms make up the second type, and in each case can be combined to produce covariant
derivatives of tensor quantities, using for example the fact that for any antisymmetric tensor field
$A^{\mu\nu}$
\begin{equation}
\frac{1}{\widehat{\sqrt{g}}}\partial_\nu(\widehat{\sqrt{g}}A^{\mu\nu})=\nabla_\nu
A^{\mu\nu}+T^\lambda_{\nu\lambda}A^{\mu\nu}+\frac{1}{2}T^{\mu}_{\lambda\nu}A^{\lambda\nu},
\end{equation}
and similar extensions including the spin connection for tensors with internal indices. The combinations
of derivatives and spin connections that appear are in fact a covariant form of the curl (the covariant
exterior derivative), in view of the antisymmetry of the tensors $F_{\mu\nu}$, $T^\alpha_{\mu\nu}$, and
$R_{\mu\nu\hphantom{a}b}^{\hphantom{\mu\nu}a}$ in $\mu$ and $\nu$. We refer to such terms as bulk
magnetization currents, in analogy with Eqs.\ (\ref{magcurrents}). [This is not entirely appropriate for
all the components, because for example the field strength $F_{\mu\nu}$ can appear in the familiar
combinations ${\bf E}^2$ and $B^2$ (for electric and magnetic fields), with different coefficients, and
the former is related to electric polarization, not magnetization. However, we will be interested in
the spacelike components and zero frequency, and then the term magnetization is
appropriate for the terms we obtain, so for simplicity we will use it for all the terms.] We identify
the covariant form of the bulk (number) magnetization as
\begin{equation}
m^{\mu\nu}_{\mathrm{b}}=\frac{1}{\widehat{\sqrt{g}}}\frac{\delta S^{\mathrm{loc}}}{\delta F_{\mu\nu}}.
\end{equation}
The covariant bulk ``energy-momentum magnetization'' is
\begin{equation}
m^{\mathrm{EM},\mu\nu}_{\mathrm{b}\hphantom{,\mu\nu}\alpha}=-\frac{1}{\widehat{\sqrt{g}}}
\frac{\delta S^{\mathrm{loc}}}{\delta T^\alpha_{\mu\nu}},
\end{equation}
the $\alpha=0$ component of which can be identified with a covariant version of the energy magnetization,
\begin{equation}
m^{\mathrm{E},\mu\nu}_{\mathrm{b}}\equiv m^{\mathrm{EM},\mu\nu}_{\mathrm{b}\hphantom{M,\mu\nu}0},
\end{equation}
and the $\alpha=a$ components are a ``momentum magnetization,''
\begin{equation}
m^{\mathrm{M},\mu\nu}_{\mathrm{b}\hphantom{,\mu\nu}a}\equiv
m^{\mathrm{EM},\mu\nu}_{\mathrm{b}\hphantom{M,\mu\nu}a}.
\end{equation}
We will refer to the resulting contribution of the first two to the currents as number or energy
magnetization currents, respectively, while for the contribution of the last to the stress tensor we refer
to it as the ``magnetization stress''. We include for completeness the bulk ``spin magnetization''
\begin{equation}
m^{\mathrm{S},\mu\lambda\hphantom{a}b}_{\mathrm{b}\hphantom{,\mu\lambda}a}=\frac{1}{\widehat{\sqrt{g}}}
\frac{\delta S^{\mathrm{loc}}}{\delta R_{\mu\lambda\hphantom{a}b}^{\hphantom{\mu\lambda}a}},
\end{equation}
although we will not need it in this work.

We can point out here that the momentum magnetization appears in both the contributions to the
momentum density and stress tensor, and the spin current. We will see that this is directly relevant to the
issue of so-called torsional Hall viscosity, and its relation with the spin density.

When we consider the contribution of such magnetization currents
to the total current flowing through a section of the system (the transport current) we must integrate
them along a hypersurface. Then we also pick up corresponding $\delta$-function current contributions on
the boundary, which arise because of the boundary term when one integrates by parts to obtain the curl
form of the bulk magnetization currents (all terms in the action are assumed to
vanish outside the boundary). Consequently, as in textbook electrodynamics of media,
the magnetization currents give no net contribution to the transport current. On the other hand, when we
look at contributions of the Chern-Simons-type terms to the currents, by construction we
\emph{cannot} find contributions that can be written as covariant derivatives of covariant tensors.
Because of this, the integration of these contributions to the currents across a section of the sample
necessarily give nontrivial contributions to the transport current.

\section{Induced action: first order in derivatives}\label{sec3}

For the remainder of this work, we will focus on $d=2$ dimensional systems. We require the induced action
to be consistent with spacetime reparametrization invariance, internal spatial rotation symmetry, and
electromagnetic gauge invariance, up to boundary terms. To do this, we must establish a consistent
filtration scheme on the myriad of terms that one could write down. We adopt a derivative counting scheme
in which $A_\mu$ and $e_\mu^\alpha$ are counted as zero derivatives, that is they are assigned degree
$0$. The spin connection $\omega_\mu$ is counted as one derivative in order to ensure that the
spacetime covariant derivative and the torsion tensor have well-defined degree one.

The naive derivative counting scheme above is complicated slightly by the special role played by the
background magnetic and gravitomagnetic fields. In a general curved spacetime, the scalar magnetic field
felt by the system is
\begin{equation}
B=\frac{1}{2}\epsilon^{\mu\nu\lambda}e_\mu^0F_{\nu\lambda}. \label{bdef}
\end{equation}
As noted above, we are also considering perturbations to the ``gravitomagnetic potential'' $e^0_\mu$, and,
noting the similarities to the electromagnetic potential $A_\mu$, we can consider correlation functions
in the presence of not only a background magnetic field, but also in the presence of a background
``gravitomagnetic field'' constructed from the timelike torsion,
\begin{equation}
B_\mathrm{G}=\frac{1}{2}\epsilon^{\mu\nu\lambda}e_\mu^0T^{0}_{\nu\lambda},
\end{equation}
which we expect to enter thermodynamic quantities similarly to the magnetic field $B$. Notice that this is
the same quantity, eq.\ (\ref{hypcond}) that is zero in a region if and only if spacelike hypersurfaces
exist there. In applications,
$B_\mathrm{G}$ will be set to zero at the end, but since it plays a similar thermodynamic role to the
magnetic field, we treat the two symmetrically for consistency.
The equilibrium properties of our system can be arbitrary functions of $B$ and $B_\mathrm{G}$, and to
capture this, it is necessary for us to retain terms at all orders in $B$ and $B_G$.

With this in mind, we adopt the following scheme for writing down terms in the induced action. Out of
all possible terms consistent with spacetime reparametrization invariance, internal rotation symmetry,
and $U(1)$ gauge symmetry (up to boundary terms), we retain terms to all orders in $B$ and $B_G$, and
only terms quadratic and to first order in derivatives in the other combinations of $A_\mu$,
$e_\mu^\alpha$, and $\omega_\mu$. This will leave us with all of those terms which contribute to linear
order in derivatives of $\psi=e_0^0-1$ and $A_\mu$ to the thermoelectric response functions. Although
we could dispense with higher order terms in the gravitomagnetic field $B_G$ and still fully capture
the thermoelectric response properties, we will see that interpreting our results will be made easier
by treating it symmetrically with the magnetic field $B$.

The most general induced action, consistent with the discussion above, is given by
\begin{align}
S^{\mathrm{eff}}=\int{d^3x}\,\widehat{\sqrt{g}}&\left[f(B,B_G)+
\gamma(B,B_G)\epsilon^{\mu\nu\lambda} e_\mu^{a}T_{\nu\lambda}^{a}\right.\nonumber\\
&\left.+\gammat(B,B_G)\epsilon^{\mu\nu\lambda}\epsilon_{ab}e_\mu^aT_{\nu\lambda}^b\right.\nonumber\\
&\left.+\frac{\nu}{4\pi}\epsilon^{\mu\nu\lambda} A_\mu\partial_\nu A_\lambda\right].
\label{effaction}
\end{align}
Here $f$, $\gamma$ and $\gammat$ are scalar functions of their arguments. We mention here that we could have
treated the two scalars constructed from the spacelike torsion, namely
\begin{align}
B_T&=\frac{1}{2}\epsilon^{\mu\nu\lambda}e_\mu^aT_{\nu\lambda}^a, \\
\widetilde{B}_T&=\frac{1}{2}\epsilon^{\mu\nu\lambda}\epsilon_{ab}e_\mu^aT_{\nu\lambda}^b
\end{align}
in a similar way as $B$ and $B_G$, keeping terms to all orders and including them in $f$, instead of only to
first order as we did. We have done it this way because it will be useful in the later discussion to
separate these pieces, and also in order to compare with the literature.

The coefficients in the effective
action, or their Taylor expansions in their arguments, correspond to response functions, as we will see.
While we are assuming the temperature is zero, we emphasize that if we do allow non-zero temperature,
the coefficients will have only exponentially small corrections, due to the gap in the energy spectrum in
the bulk.

We would now like to identify the functions appearing in the actions in Eq.\ (\ref{effaction}) with
certain thermodynamic properties of the system.
We start by computing the average currents Eqs.\ (\ref{avJ}-\ref{avtau}).
We find for the number current
\begin{align}
J^\mu&=\frac{\nu}{4\pi}\epsilon^{\mu\nu\lambda}F_{\nu\lambda}\nonumber \\
&+\frac{1}{\widehat{\sqrt{g}}}
\partial_\nu\left(\widehat{\sqrt{g}}\left(\frac{\partial f}{\partial B}+2B_T \frac{\partial
\gamma}
{\partial B}+2\widetilde{B}_T\frac{\partial\gammat}{\partial
B}\right)\epsilon^{\lambda\mu\nu}e_\lambda^0\right),\label{gencurrent}
\end{align}
from which we identify the bulk magnetization
\begin{equation}
m^{\mu\nu}_{\mathrm{b}}=\left(\frac{\partial f}{\partial B}+2B_T \frac{\partial \gamma}
{\partial B}+2\widetilde{B}_T\frac{\partial\gammat}{\partial B}\right)\epsilon^{\lambda\mu\nu}e_\lambda^0.
\label{genmag}
\end{equation}
For the spin current we find after a little algebra
\begin{equation}
J_{\mathrm{S}\hphantom{\mu}a}^{\hphantom{S}\mu\hphantom{a}b}=4\gamma e^{\mu}_0\epsilon^{ab}.
\label{genspin}
\end{equation}
We note that to this order in gradients, there is no spin magnetization as the curvature does not enter
into the action. Finally, for the energy-momentum-stress tensor we have
\begin{align}
-\tau^\mu&_{\hphantom{\mu}\alpha}=e^\mu_\alpha f
+\gamma\epsilon^{\mu\nu\lambda}T_{\nu\lambda}^a\delta_\alpha^a+\gammat\epsilon^{\mu\nu\lambda}
T_{\nu\lambda}^b\epsilon_ab\delta_\alpha^a\nonumber \\
&+\frac{1}{2}\epsilon^{\rho\nu\lambda}\left(\frac{\partial f}{\partial B}F_{\nu\lambda}+\frac{\partial
f}{\partial B_G}T^0_{\nu\lambda}\right)\left(\delta^\mu_\rho\delta_\alpha^0-e^\mu_\alpha e_\rho^0\right)
\nonumber \\
&+\epsilon^{\rho\nu\lambda}\left(\frac{\partial \gamma}{\partial B}F_{\nu\lambda}+\frac{\partial
\gamma}{\partial B_G}T^0_{\nu\lambda}\right)\left(\delta^\mu_\rho\delta_\alpha^0-e^\mu_\alpha
e_\rho^0\right)B_{T} \nonumber \\
&+\epsilon^{\rho\nu\lambda}\left(\frac{\partial \gammat}{\partial B}F_{\nu\lambda}+\frac{\partial
\gammat}{\partial B_G}T^0_{\nu\lambda}\right)\left(\delta^\mu_\rho\delta_\alpha^0-e^\mu_\alpha
e_\rho^0\right)\widetilde{B}_{T} \nonumber \\
&+\frac{1}{\widehat{\sqrt{g}}}\partial_\nu\left[\widehat{\sqrt{g}}\epsilon^{\rho\mu\nu}
\left(\frac{\partial f}{\partial B_G}+2B_T\frac{\partial \gamma}{\partial
B_G}+2\widetilde{B}\frac{\partial\gammat}{\partial B_G}\right)e_\rho^0\delta_\alpha^0\right]\nonumber \\
&+\frac{1}{\widehat{\sqrt{g}}}\partial_\nu\left(\epsilon^{\rho\mu\nu}\widehat{\sqrt{g}}2\gamma
e_\rho^a\delta_\alpha^a\right)+2\gamma\omega_{\nu\hphantom{a}c}^{\hphantom{\nu}a}\epsilon^{\rho\mu\nu}
e_\rho^c\delta^a_\alpha\nonumber\\
&+\frac{1}{\widehat{\sqrt{g}}}\partial_\nu\left(\epsilon^{\rho\mu\nu}\widehat{\sqrt{g}}2\gammat
e_\rho^a\epsilon_{ab}\delta_\alpha^b\right)+2\gammat\omega_{\nu\hphantom{a}b}^{\hphantom{\nu}a}
\epsilon^{\rho\mu\nu}
e_\rho^c\epsilon_{ac}\delta^b_\alpha.
\label{genstress}
\end{align}
The expression for the energy-momentum-stress tensor is covariant, despite its appearance (compare the
discussion in the previous section). In it, we identify the energy-momentum magnetization
\begin{align}
m^{\mathrm{EM},\mu\nu}_{\mathrm{b}\hphantom{M,\mu\nu}\alpha}=-\epsilon^{\rho\mu\nu}
&\left[\left(\frac{\partial f}{\partial B_G}+2B_T\frac{\partial\gamma}{\partial
B_G}+2\widetilde{B}_T\frac{\partial\gammat}{\partial
B_G}\right)e_\rho^0\delta_\alpha^0\right.\nonumber \\
&\left.\vphantom{\left(\frac{a}{b}\right)}+2\gamma e_\rho^a\delta_\alpha^a+2\gammat e_\rho^a\epsilon_{ab}\delta_\alpha^b\right].
\label{genemag}
\end{align}

To get a feeling for the meaning of these functions, we proceed to evaluate the currents in the absence
of any perturbations. That is, we set $e=Id$, $\omega=0$, which implies in particular that $B=B_0=F_{12}$,
$B_G=0$. We also take $B$ to be uniform in space. In this case, we find for the number current in either
ensemble
\begin{equation}
J^{\mu}(e=Id,\omega=0)=\frac{\nu B_0}{2\pi}\delta^\mu_0,
\end{equation}
allowing us to identify the unperturbed expectation of the number density
\begin{equation}
\overline{n}\equiv\frac{\nu B_0}{2\pi}.
\end{equation}
Similarly, we have for the spin current
\begin{equation}
J_{\mathrm{S}\hphantom{\mu}a}^{\hphantom{S}\mu\hphantom{a}b}(e=Id,\omega=0)=
4\gamma(B_0,0)\delta^\mu_0\epsilon^{ab},
\end{equation}
from which we can identify the unperturbed spin density
\begin{equation}
\rho_{\mathrm{S},0}=4\gamma(B_0,0).
\end{equation}
For the energy-momentum-stress tensor in flat spacetime we have
\begin{equation}
\tau^{\mu}_{\hphantom{\mu}\alpha}=-\delta^\mu_\alpha\left(f(B_0,0)-\frac{\partial f}{\partial
B}(B_0,0)B_0\right)-\delta^\mu_0\delta_\alpha^0\frac{\partial f}{\partial B}(B_0,0)B_0,
\end{equation}
which allows us to identify $-f(B_0,0)$ as the unperturbed energy density (as is clear from the effective
action itself, as we are using the canonical ensemble). The unperturbed internal pressure
\cite{Cooper1997,Bradlyn2012} is
\begin{equation}
p_{\mathrm{int},0}\equiv f(B_0,0)-\frac{\partial f}{\partial B}(B_0,0)B_0.
\end{equation}

Lastly, we examine the unperturbed magnetizations for $e=Id$, $\omega=0$, as these expressions will
prove useful later. We find for the bulk number magnetization
\begin{equation}
m^{\mu\nu}_{\mathrm{b},0}=\frac{\partial f}{\partial B}(B_0,0)\epsilon^{0\mu\nu};
\end{equation}
for the bulk energy magnetization,
\begin{equation}
m^{\mathrm{E},\mu\nu}_{\mathrm{b},0\hphantom{\nu}}=-\epsilon^{0\mu\nu}\frac{\partial f}{\partial
B_G}(B_0,0);
\end{equation}
and for the bulk momentum magnetization
\begin{equation}
m^{\mathrm{M},\mu\nu}_{\mathrm{b},0\hphantom{\nu}a}=2\epsilon^{b\mu\nu}\gammat(B_0,0)\epsilon_{ab}
-2\epsilon^{a\mu\nu}\gamma(B_0,0).
\end{equation}

\section{Linear Response from the Induced Action}\label{respsec}

\subsection{General considerations}

With all this formalism established, we now wish to examine the response of the average currents to the
external fields to linear order. Before we proceed to expand the expressions Eqs.\
(\ref{gencurrent}-\ref{genstress}) in the external fields, we must connect our
currents with those in the statistical physics literature. To do so, we must make contact with the standard
view of the perturbing fields $\delta e$ and $\omega$ as externally applied
fields\cite{Luttinger1964,Cooper1997}.

While what we have done up to now is valid in any system of coordinates, we must remember that a
physical measurement is performed using a fixed choice of coordinates $x^\mu$
(the lab coordinate system, if you will). We would like to interpret the vielbeins $e_\mu^a(x)$ as
externally applied
fields in this given coordinate system. If they were held fixed, then this does not cause any issues,
but because we wish to vary them and study the response to perturbations in them, it is necessary to be
careful about the following point. Given a conserved vector field $K^\mu$ (such as the conserved number
current $J^\mu$, and others), we identify the experimentally relevant current by considering
the flux of $K^\mu$ through a surface that is {\em fixed} when the perturbing $\delta e_\mu^\alpha$. It is
of paramount importance to maintain conservation of $K^\mu$; so by considering integrals of the two-form
\begin{equation}
K^\mu\epsilon_{\mu\nu\lambda}dx^\nu dx^\lambda \nonumber
\end{equation}
across an infinitesimal hypersurface, which as we saw in Sec.\ \ref{sec2} requires no addition vielbein
factors under the integral, we see that it is actually $K^\mu\epsilon_{\mu\nu\lambda}$ that we
should utilize. Extracting the coordinate-transformation invariant $\check{\epsilon}_{\mu\nu\lambda}$
symbol, we see that the physically-meaningful quantity is the tensor \emph{density}
\begin{equation}
\widehat{K}^\mu=\widehat{\sqrt{g}}K^\mu.
\end{equation}
The practical effect of this is that when we look at the change in $\widehat{K}^\mu$ to linear order in
perturbations of the vielbeins, there is an additional term compared with what one obtains using $K^\mu$.
(In a microscopic linear response calculation, these show up as ``contact terms'', that is contributions
to the response that are given by an expectation value of some operator at a single time, rather like
the familiar diamagnetic term in conductivity response.)

As a example of how this makes contact with the literature, let us revisit the microscopic number
current computed in Sec.\ \ref{sec2}. In the presence of nontrivial $e_\mu^\alpha$, the number current
computed from Eq.\ (\ref{microaction}) is
\begin{equation}
J^\mu=e^\mu_0\varphi^\dag \varphi-\frac{i}{2m}e^\mu_ae^\nu_a\left(\varphi^\dag
D_\nu\varphi-(D_\nu\varphi)^\dag\varphi\right),
\end{equation}
whereas the number current density is given by
\begin{equation}
\widehat{J}^\mu=\widehat{\sqrt{g}}\left(e^\mu_0\varphi^\dag
\varphi-\frac{i}{2m}e^\mu_ae^\nu_a\left(\varphi^\dag D_\nu\varphi-(D_\nu\varphi)^\dag\varphi\right)\right).
\end{equation}
Let us examine this in the case of Luttinger's gravitational perturbation in otherwise flat space, setting
\begin{align}
e_\mu^0&=\delta_\mu^0\left(1+\psi\right), \nonumber \\
e_\mu^a&=\delta_\mu^a. \label{luttinger}
\end{align}
We then find that
\begin{equation}
\widehat{J}^\mu=\delta^\mu_0\varphi^\dag\varphi-\left(1+\psi\right)\delta^\mu_i\frac{i}{2m}
\left(\varphi^\dag D_i\varphi-(D_i\varphi)^\dag\varphi\right),
\end{equation}
in agreement with the form of the current operator in the presence of the background gravitational
field presented in Refs. \onlinecite{Luttinger1964,Cooper1997}.

Similar considerations hold for the energy-momentum-stress energy tensor $\tau^\mu_{\hphantom{\mu}\alpha}$.
In that case we must also pay attention to the second (lower) index. The physical response corresponds to
the tensor density with the second index converted to an ambient spacetime index in the same lab coordinate
system. For example, in the case
of Luttinger's perturbation, this corresponds to the Hamiltonian being the generator of translations
along the vector field $\partial/\partial x^{\mu=0}$ rather than along $e^\mu_0\partial/\partial x^\mu$.
Altogether, we must consider the response of the energy-momentum-stress tensor density
$\widehat{\tau}^\mu_{\hphantom{\mu}\nu}$ to perturbations.
As an illustration, if we consider the energy density $\widehat{\tau}^0_{\hphantom{0}\nu=0}$ computed
from the microscopic action Eq.\ (\ref{microaction}) in the presence of Luttinger's gravitational
perturbation Eq.\ (\ref{luttinger}) we find
\begin{equation}
\widehat{\tau}^0_{\hphantom{0}\nu=0}=(1+\psi)\frac{1}{2m}\left(D_i\varphi\right)^\dag D_i\varphi,
\label{needaname}
\end{equation}
consistent with the energy density operator used in Refs. \onlinecite{Luttinger1964,Cooper1997} for
calculating thermal transport coefficients.

Following this discussion, the current densities we wish to consider are, from Eqs.\
(\ref{gencurrent}-\ref{genstress})
\begin{align}
\widehat{J}^\mu=&\frac{\nu}{4\pi}\widehat{\epsilon}^{\mu\nu\lambda}F_{\nu\lambda}+
\partial_\nu\widehat{m}^{\mu\nu}_{\mathrm{b}},\label{gencurrentdensity}\\
\widehat{\tau}^{\mu}_{\hphantom{\mu}\nu}&=-\widehat{\sqrt{g}}\delta^\mu_\nu
f-\gamma\widehat{\epsilon}^{\mu\rho\lambda}
T_{\rho\lambda}^ae_\nu^a
-\gammat\widehat{\epsilon}^{\mu\rho\lambda}T_{\rho\lambda}^b\epsilon_{ab}e_\nu^a\nonumber \\
&-\frac{1}{2}\widehat{\epsilon}^{\rho\sigma\lambda}\left(\frac{\partial f}{\partial
B}F_{\sigma\lambda}+\frac{\partial f}{\partial
B_G}T^0_{\sigma\lambda}\right)\left(e_\nu^0\delta^\mu_\rho-e_\rho^0\delta^\mu_\nu\right) \nonumber \\
&-B_T\widehat{\epsilon}^{\rho\sigma\lambda}\left(\frac{\partial \gamma}{\partial
B}F_{\sigma\lambda}+\frac{\partial \gamma}{\partial
B_G}T^0_{\sigma\lambda}\right)\left(e_\nu^0\delta^\mu_\rho-e_\rho^0\delta^\mu_\nu\right) \nonumber \\
&-\widetilde{B}_T\widehat{\epsilon}^{\rho\sigma\lambda}\left(\frac{\partial \gammat}{\partial
B}F_{\sigma\lambda}+\frac{\partial \gammat}{\partial
B_G}T^0_{\sigma\lambda}\right)\left(e_\nu^0\delta^\mu_\rho-e_\rho^0\delta^\mu_\nu\right) \nonumber \\
&-e_\nu^\alpha\partial_\sigma
\widehat{m}^{\mathrm{EM}\mu\sigma}_{\mathrm{b}\hphantom{E\mu\nu}\alpha}-\omega_{\sigma\hphantom{a}c}
^{\hphantom{\nu}a}\widehat{m}^{\mathrm{M}\mu\sigma}_{\mathrm{b}\hphantom{,\mu\nu}a}e_\nu^c.
\label{genstressdensity}
\end{align}
Here we may mention that because of the use of the current densities, the Hall conductivity
(the first term in the current density response) comes out as $\nu/(2\pi)$, which is quantized,
times the coordinate-independent $\widehat{\epsilon}^{ij}$, showing quantization with no need to extract
a factor involving the vielbeins.

\subsection{Thermoelectric Response}
\label{sec:thermo}

We first consider number and energy current density response to Luttinger's $\psi=e_0^0-1$ and an electric
potential $\phi=-A_0$ to obtain the full set of electric, thermal, and cross conductivities.
We assume that $F_{12}=B_0$ is independent of space and time coordinates. A convenient fact about this
choice of perturbing fields is that $B=B_0$ and $B_G=0$, independent of $\psi$. In particular, this ensures
that
\begin{equation}
\partial_\mu\frac{\partial f}{\partial B}=\partial_\mu\frac{\partial f}{\partial B_G}=0
\end{equation}
in the Luttinger case. Using this fact, we can expand the number and energy current densities, Eqs.\
(\ref{gencurrentdensity}) and (\ref{genstressdensity}), to first order in $\phi$ and $\psi$ to find
\begin{align}
\widehat{J}^i&=-\frac{\nu}{2\pi}\widehat{\epsilon}^{ij}\partial_j\phi+\partial_j
\left[\widehat{m}^{ij}_{\mathrm{b},0}\left(1+\psi\right)\right] ,\label{currentresp1}\\
\widehat{J}_E^i&\equiv\widehat{\tau}^i_{\hphantom{i}\nu=0}=\partial_j\left(\widehat{m}^{ij}_{\mathrm{b},0}
\phi\right)+\partial_j\left[\widehat{m}^{\mathrm{E},ij}_{\mathrm{b},0\hphantom{\nu}}\left(1+2\psi\right)
\right]. \label{currentresp2}
\end{align}
Comparing with eqs.\ (\ref{magcurrents}), (\ref{magperts}), we see that apart from the Hall conductivity
term in $\widehat{J}^\mu$, the terms are precisely the magnetization contributions, though in the present
case they result from the bulk only.
In particular, this allows us to identify the kinetic coefficients $L^{(n)}$ (which obey Onsager
reciprocity provided $\widehat{m}^{ij}_{\mathrm{b},0}$ is an odd function of $B$)
\begin{align}
L^{(1)}_{ij}&=\frac{\nu}{2\pi}\widehat{\epsilon}^{ij} ,\label{l1}\\
L^{(2)}_{ij}&=L^{(3)}_{ij}=-\widehat{m}^{ij}_{\mathrm{b},0}=\frac{\partial f}{\partial
B}\widehat{\epsilon}^{ij} ,\label{l23}\\
L^{(4)}_{ij}&=-2\widehat{m}^{\mathrm{E},ij}_{\mathrm{b},0\hphantom{\nu}}=2\frac{\partial f}{\partial
B_G}\widehat{\epsilon}^{ij} \label{l4}.
\end{align}
This is one of our main results, and requires further discussion. The bulk magnetization currents are
equilibrium effects, and because of boundary contributions to the current from the same terms in the
action do not contribute to the net current across
any section of the system. Neither does $L^{(4)}$ bear any particular relation to the central charge $c$.

Comparing with the work of CHR, in their case the magnetization includes edge effects, and these are not
just the contributions that relate to the bulk magnetization (the latter is temperature independent, up
to exponentially small corrections, while there are thermal edge currents of order $T^2$). Moreover,
the transport current densities, which correspond to the net current through a section across the sample,
were declared to be due to bulk transport current density with no contribution located on the edge,
{\em by definition}. When this is done, the effect of thermal excitation at the edge that produces the
thermal Hall conductivity (related to the central charge) is reassigned as a bulk effect. At the same time
the bulk magnetization effects are canceled in the transport current densities by the corresponding part
of the edge currents as we have seen. (As they emphasize, the actual local current density in the bulk,
which is what we have studied, is not the same as the transport current density.) In equations, their
prescription for the transport coefficients is given above in Eq.\ (\ref{transportresp2}). (At this stage,
they are written for the response of the heat, not energy, current density to perturbations that couple
to number and heat, not energy.) If we use our results along with the known $O(T^2)$ edge contribution
to the energy magnetization, we finally obtain
\begin{align}
N^{(1)}_{ij}&=\frac{\nu}{2\pi}\widehat{\epsilon}^{ij}, \nonumber \\
N^{(2)}_{ij}&=N^{(3)}_{ij}=0, \nonumber \\
N^{(4)}_{ij}&=\frac{\pi c}{6}T^2\widehat{\epsilon}^{ij}.
\end{align}
These results are checked explicitly for a non-interacting integer quantum Hall system in Appendix
\ref{app}.

We see then that, excepting the Hall current, the thermoelectric transport currents are due solely to
edge effects, and flow along the edge, even if gravitational background fields are included.
(In the case of the Hall number current density response, in the
more general situation in which there is a bulk electric field as well as a chemical potential
gradient, there are contributions from both the bulk and the edge, such that the net current through
a section is proportional to the change in the electrochemical potential across the sample, which is
the statement of the quantized Hall effect. This was well understood in the 1980s,
but is the subject of frequent misstatements at present.) The confusion that exists in the literature
concerning the thermal Hall conductivity arises from the aforementioned fact that the total
magnetization densities, and thus the transport current densities, are typically defined by fiat to include
effects from the edge. We have shown here that bulk thermoelectric response is independent of these
edge contributions. This result should be contrasted with the recent claims of Ref.\
\onlinecite{Shitade2013}.

\subsection{Stress Response}
\label{sec:stress}

Finally we consider the response of the stress tensor to time-varying spatial perturbations $\delta
e_i^a(t)$ of the vielbeins, once again with a spatially uniform and time independent electromagnetic
field strength $F_{12}=B_0$.

Expanding Eq.\ (\ref{genstressdensity}) to linear order in the perturbing fields, we find
\begin{align}
-\widehat{\tau}^{i}_{\hphantom{i}j}=&\delta^i_j\left(p_{\mathrm{int},0}+\left(p_{\mathrm{int},0}
+B_0^2\frac{\partial^2f}{\partial B^2}\right)\mathrm{tr}(\delta e_k^a)\right) \nonumber \\
&+B_0\frac{\partial\gamma}{\partial
B}\widehat{\epsilon}^{a\nu\lambda}T_{\nu\lambda}^a\delta^i_j+\gamma\widehat{\epsilon}^{i\rho\lambda}
T^{a}_{\rho\lambda}\delta_j^a \nonumber \\
&+2\delta_j^a\widehat{\epsilon}^{i\sigma\rho}\left(\partial_\sigma\left(\gamma
e_\rho^a\right)+\gamma\omega_{\sigma\hphantom{a}b}^{\hphantom{\sigma}a}e_\rho^b\right).
\end{align}
All terms proportional to $\gammat$ cancel. Using the structure equation Eq.\ (\ref{cartan}), we can
combine the second, third, and fourth terms to obtain, when $\omega=0$,
\begin{align}
-\widehat{\tau}^{i}_{\hphantom{i}j}=&\delta^i_j\left(p_{\mathrm{int},0}
+\left(p_{\mathrm{int},0}+B_0^2\frac{\partial^2f}{\partial B^2}\right)\mathrm{tr}(\delta
e_k^a)\right)\nonumber \\
&+2\gamma\left(\delta_j^a\epsilon^i_{\hphantom{i}\ell}-\delta^i_\ell\epsilon^{a}_{\hphantom{a}j}
\right)\partial_0e^a_\ell\nonumber \\
&+\left(\gamma-B_0\frac{\partial\gamma}{\partial
B}\right)\left(\epsilon^i_{\hphantom{i}j}\delta^\ell_a-\delta^i_j\epsilon^{\ell a}\right)\partial_0e^a_\ell.
\end{align}
The first term is what is expected for the response to dilations\cite{Bradlyn2012}, and allows us to
identify the inverse internal compressibility
\begin{equation}
\kappa^{-1}_\mathrm{int}=-B^2\frac{\partial^2f}{\partial B^2}.
\end{equation}
The second arises from the spacelike torsion term with coefficient $\gamma$, and gives what has been
called ``torsional Hall viscosity''\cite{Hughes2013} and is in fact equal to one-half the unperturbed
spin density as in Refs.\ \onlinecite{Read2009,Read2011}.
The final term breaks the symmetry of the stress tensor, and is necessary to ensure that the continuity
equation Eq.\ (\ref{stresscont}) is satisfied.

Even though we will not analyze all possible two-derivative terms in this paper, we will also examine the
contribution of the first Wen-Zee term
\begin{equation}
S^{\mathrm{WZ1}}=\frac{\nu\mathcal{S}}{4\pi}\int{d^3x}\widehat{\sqrt{g}}\epsilon^{\mu\nu\lambda}
\omega_\mu\partial_\nu A_\lambda \label{wz1}
\end{equation}
to the stress tensor. Here $\mathcal{S}$ is the shift\cite{Wen1992}. While of degree two in our counting
scheme, it should be kept here because we keep
the field strength $B_0$ as if it were of degree zero. We find, trivially, that this term contributes
nothing to the stress, as the spin connection and vector potential are independent of the vielbeins. This
is contrary to our expectation that the first Wen-Zee term furnish a Hall viscosity with a coefficient
containing $\mathcal{S}/4$ \cite{Read2009b,Hoyos2012}. It does however contribute to the spin
current a term $\frac{\nu\mathcal{S}}{4\pi}\epsilon^{\mu\nu\lambda}\partial_\nu A_\lambda$.

To make sense of these results, we now argue that Eq.\ (\ref{genstressdensity}) is not the physical
stress tensor corresponding to momentum transport, as Eq.\ (\ref{antisym}) shows that it is not symmetric
even in the absence of torsion. For various reasons\cite{Martin,Weinberg}, it is preferable to use a
symmetric stress tensor. This is accomplished with the Belinfante ``improved'' energy-momentum-stress
tensor density. We obtain this by changing our view of which variables are independent in the description
of the spacetime geometry. Instead of $e_\mu^\alpha$ and $\omega_\mu$, from which the Christoffel symbols
and torsion were derived using covariant constancy of the vielbein, we will now change to using
$e_\mu^\alpha$ (or the corresponding tensors with upper and lower indices interchanged) and the reduced
torsion $\Tt^a_{\mu\nu}$ defined in Appendix \ref{appB} as the independent variables (the reduced torsion
has the same number of independent components as the spin connection). We show in Appendix \ref{appB}
that the Christoffel symbols, spin connection, and spacelike torsion can be expressed in terms of these
(we already know that the timelike torsion can be). Then we make the definition
\begin{equation}
\tau^{\hphantom{\mathrm{B}}\mu}_{\mathrm{B}\hphantom{\mu}\alpha}=-\frac{1}{\widehat{\sqrt{g}}}
\left(\frac{\delta S^{\mathrm{eff}}}{\delta e_\mu^\alpha}\right)_{A,\Tt^a},
\label{genBeltens}
\end{equation}
and call this the generalized Belinfante energy-momentum-stress tensor, because it resembles the
Belinfant improvement procedure (which we don't describe, but it involves adding derivatives of the spin
current to the energy-momentum-stress tensor), and ``generalized'' because we include torsion.
Further details are in Appendix \ref{appB}; we note here only that the space components are symmetric in
the absence of reduced torsion. In general, the change in definition has the consequence that the Belinfante
energy-momentum-stress tensor differs from $\tau^\mu_\alpha$ by a change in the momentum magnetization
from the locally-invariant terms in $S^{\mathrm{eff}}$, and by the appearance
of terms coming from the Riemann tensor in the effective action, which however we don't have in our
first-order $S^{\mathrm{eff}}$.
Clearly we should use this definition throughout, including for the
thermoelectric responses in the previous section. However, for those responses, the change in definition
makes no difference. Finally, we will continue to refer to the spin current with the same definition as
before for convenience, however the formalism leads us to introduce another field which takes the place
of the spin current in some expressions, which is
\begin{equation}
\theta^{\mu\nu}_{\hphantom{\mu\nu}a}=\frac{1}{\widehat{\sqrt{g}}}\left(\frac{\delta S^{\mathrm{eff}}}
{\delta \Tt^{a}_{\mu\nu}}\right)_{A,e^\alpha},
\label{anaspincurr}
\end{equation}
which in fact is exactly the part of the momentum magnetization removed in this construction, and so
appeared as a term in the spin current (up to a vielbein factor).

Using our first-order effective action, we find for the Belinfante energy-momentum-stress tensor density
\begin{align}
\widehat{\tau}^{\hphantom{\mathrm{B}}\mu}_{\mathrm{B}\hphantom{\mu}\nu}&=-\widehat{\sqrt{g}}\delta^\mu_\nu
f-\gamma\widehat{\epsilon}^{\mu\rho\lambda}\Tt_{\rho\lambda}^ae_\nu^a
-2\gammat\widehat{\epsilon}^{\mu\rho\lambda}(\partial_\rho e_\lambda^b)\epsilon_{ab}e_\nu^a\nonumber \\
&-\frac{1}{2}\widehat{\epsilon}^{\rho\sigma\lambda}\left(\frac{\partial f}{\partial
B}F_{\sigma\lambda}+\frac{\partial f}{\partial
B_G}T^0_{\sigma\lambda}\right)\left(e_\nu^0\delta^\mu_\rho-e_\rho^0\delta^\mu_\nu\right) \nonumber \\
&-B_T\widehat{\epsilon}^{\rho\sigma\lambda}\left(\frac{\partial \gamma}{\partial
B}F_{\sigma\lambda}+\frac{\partial \gamma}{\partial
B_G}T^0_{\sigma\lambda}\right)\left(e_\nu^0\delta^\mu_\rho-e_\rho^0\delta^\mu_\nu\right) \nonumber \\
&-\widetilde{B}_T\widehat{\epsilon}^{\rho\sigma\lambda}\left(\frac{\partial \gammat}{\partial
B}F_{\sigma\lambda}+\frac{\partial \gammat}{\partial
B_G}T^0_{\sigma\lambda}\right)\left(e_\nu^0\delta^\mu_\rho-e_\rho^0\delta^\mu_\nu\right) \nonumber \\
&-e_\nu^b\partial_\nu\left(\epsilon^{\rho\mu\nu}\widehat{\sqrt{g}}2\gammat
e_\rho^a\epsilon_{ab}\right)-e_\nu^0\partial_\sigma\widehat{m}^{\mathrm{E}\mu\sigma}_{\mathrm{b}
\hphantom{\mathrm{E}\mu\nu}},
\end{align}
which, as noted above, differs from Eq.\ (\ref{genstressdensity}) in that it receives no
magnetization stress contribution from the reduced torsion.

Expanding to linear order in the perturbing fields $\delta e_i^a(t)$ and $\Tt_{\mu\nu}^a$, we find
\begin{align}
-\widehat{\tau}^{\hphantom{\mathrm{B}}i}_{\mathrm{B}\hphantom{i}j}=&\delta^i_j\left(p_{\mathrm{int},0}
+\left(p_{\mathrm{int},0}+B_0^2\frac{\partial^2f}{\partial B^2}\right)\mathrm{tr}(\delta e_k^a)\right)
\nonumber \\
&+B_0\frac{\partial\gamma}{\partial
B}\epsilon^{a\nu\lambda}\Tt_{\nu\lambda}^a\delta^i_j+\gamma\epsilon^{i\rho\lambda}\Tt^{a}_{\rho\lambda}
\delta_j^a.
\end{align}
The first term here is unchanged. From the tensor structure of the second term, we see that it gives a
change in pressure in the presence of background reduced spacelike torsion. The last term breaks the
symmetry of the stress tensor in the presence of reduced torsion, which is necessary given the
symmetrization condition Eq.\ (\ref{belinfantesym}) derived in Appendix \ref{appB}.
Some intuition for these two terms can be gleaned from the fact that they can be re-expressed as
\begin{equation}
B_0\frac{\partial\gamma}{\partial
B}\epsilon^{a\nu\lambda}\Tt_{\nu\lambda}^a\delta^i_j+\gamma\epsilon^{i\rho\lambda}\Tt^{a}_{\rho\lambda}
\delta_j^a
=\frac{1}{2}e_j^\alpha
\Tt^a_{\rho\lambda}\frac{\delta\widehat{\theta}^{\rho\lambda}_{\hphantom{\rho\lambda}a}}{\delta
e_i^\alpha}\label{spinform}
\end{equation}
expanded to linear order in the external fields. We thus see that these two contributions to the Belinfante
tensor correspond to the change in momentum magnetization density due to strain.
The stress tensor does not receive a contribution of the form of the ``torsional Hall viscosity''
mentioned above. A similar effect was noted in Ref.\
\onlinecite{Hoyos2014} for the relativistic case. If we define viscosity as the response to
$\partial_0e^a_i$ at zero reduced spacelike torsion, then we obtain no viscosity terms at all from the
first-order action.

In addition, we now make an important point: the minimally-coupled microscopic matter
action in Sec.\ \ref{sec:matter} does not feel reduced torsion at all in the case of non-interacting
particles, or
of particles with a $\delta$-function interaction, nor for general potential interactions at least when
$B_G=0$ (so spacelike hypersurface exist). In all these cases, the microscopic action depends
only on the vielbein and vector potential, not on the reduced torsion. Consequently (see Eqs.\
(\ref{redtorsscal1}), (\ref{redtorsscal2}) in the Appendix), for all such cases
the coefficient function $\gamma$ in the first-order effective action is zero, at least for $B_G=0$:
\be
\gamma(B,0)=0,
\ee
while $\widetilde{\gamma}$ does not have to vanish.
Hence in these cases the unusual contributions
to the stress response are simply absent, and both the current $\theta^{\mu\nu}_a$ and the spin current
$J^{\hphantom{S}\mu\hphantom{a}b}_{S\hphantom{\mu}a}$ resulting from the first-order action are zero.

Finally, our construction of the Belinfante stress tensor allows us to see how the first Wen-Zee term
furnishes a Hall viscosity even in the absence of reduced torsion (we already saw
that it produces an addition to the spin current). Eq.\ (\ref{spincon}) allows us to express
the first Wen-Zee term solely in terms of the electromagnetic field strength, the reduced torsion, and
the spacelike vielbeins. Modulo reduced torsion terms that are locally invariant, which we will drop,
the first Wen-Zee term Eq.\ (\ref{wz1}) becomes
\begin{equation}
S^{\mathrm{WZ}1}\sim\frac{\nu\mathcal{S}}{16\pi}\int{d^3x\widehat{\sqrt{g}}\epsilon^{\mu\nu\lambda}
\epsilon^{ab}F_{\mu\nu}\left(e^\rho_ae^\sigma_b\partial_\sigma h_{\rho\lambda}+e^\rho_a\partial_\lambda
e_\rho^b\right)}.
\end{equation}
Computing the contribution of this term in the action to the Belinfante stress tensor to linear order in
the perturbations yields an additional contribution
\begin{equation}
\Delta\widehat{\tau}^{\hphantom{\mathrm{B}}i}_{\mathrm{B}\hphantom{i}j}=\frac{1}{4}\overline{n}
\mathcal{S}\left(\epsilon^{i\ell}\delta_{kj}-\epsilon^{kj}\delta_{i\ell}\right)\partial_0e^\ell_k.
\end{equation}
As expected, this has the form of a Hall viscosity, with known coefficient\cite{Read2009,Read2011}
\begin{equation}
\eta^\mathrm{H}=\frac{1}{4}\overline{n}\mathcal{S}.
\end{equation}
Because the Wen-Zee term is not locally invariant, this contribution is not a magnetization stress, and the
coefficient $\nu {\cal S}$ is robust against perturbations of the model that maintain the gap and preserve
all symmetries. This confirms that previous Berry phase\cite{Read2009,Read2011} and linear
response\cite{Bradlyn2012} calculations of the Hall viscosity yielded a true transport coefficient.
The locally-invariant contributions which we have ignored in this analysis only add to
the expression for $\theta^{\mu\nu}_{\hphantom{\mu\nu}a}$ appearing in Eq.\ (\ref{spinform}). They
contribute nothing to the stress tensor when the reduced torsion is zero, and cannot arise in the
minimally-coupled models when $B_G$ is zero. In the latter case, for the
action we have considered, the spin current is due solely to the Wen-Zee term as well, and the
relation of the Hall viscosity with the spin density\cite{Read2009,Read2011}
$\overline{n}\overline{s}=\overline{n}{\cal S}/2$ is found also (we expect this relation to
be maintained when other second- or higher-order terms are included as well). We also note that the Wen-Zee
term as above does not contribute to the thermoelectric transport.

\section{Conclusion}

We have found a low-energy induced bulk action for transport in gapped topological phases by
allowing the spacetime geometry to include timelike and spacelike torsion as well as curvature. \
From this, we derived the bulk thermoelectric transport
coefficients, and showed that a gapped bulk cannot contribute to thermal conductivity or thermopower, up to
exponentially small corrections in temperature. We examined the stress tensor, and showed
that any torsional Hall viscosity
drops out in the appropriate Belinfante improved tensor, leaving the Hall viscosity that is related to
the orbital spin density.

A similar approach can be taken for other terms in the action that are higher than first order in
derivatives in our counting scheme. These will not contribute directly to transport, but we expect
the central charge to appear as a coefficient. We defer the treatment of these terms to a future work.

\acknowledgements

B.B. would like to thank A. Gromov and M. Goldstein for fruitful discussions, as well as W. Goldberger and
D. Poland for their insight into the nuances of Riemannian geometry. NR is grateful to W. Goldberger and
D. Son for helpful discussions. The work of B.B. and N.R. was supported by NSF grant No.\ DMR-1005895.

\appendix

\section{Generalized Belinfante Construction}\label{appB}

In this appendix, we generalize the Belinfante construction of a symmetric stress tensor to situations
in which spacetime has torsion. Our guiding principle is that we demand that the
stress tensor correspond as closely as possible with a variation of the action with respect to the
(degenerate) spatial metric
\begin{equation}
h_{\mu\nu}=e_\mu^ae_\nu^a
\end{equation}
rather than to a variation of the action with respect to the vielbein with the spin connection held
fixed. If the spin connection can be expressed in terms of the vielbein and torsion, and if these are
independent (if there are no relations between torsion and vielbeins), then varying the vielbeins with
the torsion held fixed will produce such a stress tensor, very much in analogy with the
usual case (in particular, when torsion is absent throughout). In our case, the timelike torsion and,
as it turns out, also part of the spacelike torsion are determined by the vielbeins alone, independent of
the spin connection, so that if we desire (as we do) to have no constraints on the vielbeins, we cannot
take all components of torsion as independent, because for example they cannot all be set to zero
without introducing unwanted constraints on the vielbeins.

First we solve Eq.\ (\ref{covconsteq}) for the Christoffel symbols and the
spin connection. To this end, we note first that an immediate consequence of the covariant constancy of
the vielbein is the covariant constancy of the degenerate metric $h_{\mu\nu}$, that is
\begin{equation}
\nabla_{\mu}h_{\nu\lambda}=0=\partial_\mu h_{\nu\lambda}
-\Gamma^{\rho}_{\hphantom{\rho}\mu\nu}h_{\rho\lambda}-\Gamma^{\rho}_{\hphantom{\rho}\mu\lambda}h_{\nu\rho}.
\end{equation}
We can solve this equation for the symmetric part of the Christoffel symbols, while the antisymmetric part
is determined solely by the torsion tensor (timelike and spacelike). The result can be expressed most
simply as
\begin{align}
\Gamma^\alpha_{\hphantom{\alpha}\lambda\nu}\equiv
&e_\mu^\alpha\Gamma^{\mu}_{\hphantom{\mu}\lambda\nu}=\delta^\alpha_0\partial_\lambda e_\nu^0\nonumber \\
&+\delta^\alpha_a\left[\frac{1}{2}\eta^{ab}e^\mu_b\left(\partial_\nu h_{\lambda\mu}+\partial_\lambda
h_{\mu\nu}-\partial_\mu h_{\nu\lambda}\right)+K^a_{\lambda\nu}\right], \label{christoffels}
\end{align}
where we have introduced the contorsion tensor
\begin{equation}
K^a_{\lambda\nu}=\frac{1}{2}\left[T^a_{\lambda\nu}+\eta^{ab}\left(e^\mu_be^c_\lambda
T^c_{\mu\nu}+e^\mu_be^c_\nu T^c_{\mu\lambda}\right)\right].
\end{equation}

Next, with explicit expressions for the Christoffel symbols in hand, we wish to solve Eq.\
(\ref{covconsteq}) for the spin connection. Examination of Eq.\ (\ref{cartan}) shows that (similar to
the case of the timelike torsion) because the spin connection vanishes when either of its internal indices
are timelike, there is a part of the spacelike torsion that is independent of the
spin connection. This part can be expressed as
\begin{align}
C^{ab}&\equiv e^{\mu}_0e^\nu_c\left(\eta^{ac}T^b_{\mu\nu}+\eta^{bc}T^a_{\mu\nu}\right) \label{Cdef}\\
&=\left(e^\mu_0e^\nu_c-e^\nu_0e^\mu_c\right)\left(\eta^{ac}\partial_\mu
e_\nu^b+\eta^{bc}\partial_\mu e_\nu^a\right).
\end{align}
This allows us to define what we will call the \emph{reduced torsion}, which is purely spacelike:
\begin{equation}
\Tt^a_{\mu\nu}\equiv T^a_{\mu\nu}-\frac{1}{2}\eta_{bc}C^{ab}\left(e_\mu^0e_\nu^c-e_\nu^0e_\mu^c\right).
\label{Ttildedef}
\end{equation}
The components of the reduced torsion are not all independent; it is defined so that it yields zero when
substituted into the definition of $C^{ab}$, Eq.\ (\ref{Cdef}). This is natural: we are seeking a linear
relation between the torsion and spin connection, but the latter has $d(d-1)(d+1)/2$ independent
components, while spacelike torsion has $d^2(d+1)/2$ independent components. Taking into account the
$d(d+1)/2$ constraints from setting Eq.\ (\ref{Cdef}) to zero, we are left with $d(d-1)(d+1)/2$ independent
components of reduced spacelike torsion, the same as in the spin connection, as required.

Specializing to $2+1$ dimensions, we can now solve Eq. (\ref{covconsteq}) for the spin connection in terms
of the reduced torsion and the vielbeins to find
\begin{align}
\omega_\lambda\equiv&\frac{1}{2}\epsilon_a^{\hphantom{a}b}\omega^{\hphantom{\lambda}a}
_{\lambda\hphantom{a}b}=\frac{1}{2}\epsilon^{ab}e^\mu_ae^\nu_b\left(\partial_\nu
h_{\mu\lambda}+\frac{1}{2}e_\lambda^c\Tt^c_{\mu\nu}\right)\nonumber \\
&+\frac{1}{2}\epsilon^{ab}e^\mu_a\left(\partial_\lambda e_\mu^b+\Tt_{\mu\lambda}^b\right). \label{spincon}
\end{align}
There are similar expressions in higher dimensions. We thus see that we are free to consider the
reduced torsion, instead of the spin connection, as an independent variable along with the
vielbeins and the U($1$) vector potential $A_\mu$. We also note that
the scalars constructed from the torsion that were defined in Sec.\ \ref{sec3} can be written as
\begin{align}
B_T&=\frac{1}{2}\epsilon^{\mu\nu\lambda}e^a_\mu\Tt^a_{\nu\lambda},\label{redtorsscal1}\\
\widetilde{B}_T&=\frac{1}{2}\eta_{ab}C^{ab}.\label{redtorsscal2}
\end{align}

The preceding allows us to define the generalized Belinfante energy-momentum-stress tensor
$\tau^{\hphantom{\mathrm{B}}\mu}_{\mathrm{B}\hphantom{\mu}\alpha}$ resulting from an action $S$ as in Eq.\
(\ref{genBeltens}),
where the reduced torsion is held fixed in the functional derivative.
We claim that $\tau_B$ represents the \emph{physical} energy-momentum-stress tensor. To justify this, we
must derive the continuity equation that it satisfies. To do so, we also need $\theta^{\mu\nu}_a$ as
defined in Eq.\ (\ref{anaspincurr}); it is the analog of the spin current (which has the same number of
independent components).
Examining the variation of a general action under spacetime diffeomorphism as in Sec. \ref{sec2},
we find that the generalized Belinfante
energy-momentum-stress tensor satisfies the continuity equation (after use of the equations of motion,
if any)
\begin{align}
\nabla_\mu\tau^{\hphantom{\mathrm{B}}\mu}_{\mathrm{B}\hphantom{\mu}\lambda}
-T^\rho_{\rho\mu}\tau^{\hphantom{\mathrm{B}}\mu}_{\mathrm{B}\hphantom{\mu}\lambda}=&2\Tt^{a}_{\mu\lambda}
\left(\nabla_\nu\theta^{\mu\nu}_{\hphantom{\mu\nu}a}-T^{\rho}_{\nu\rho}\theta^{\mu\nu}_{\hphantom{\mu\nu}a}
\right)\nonumber \\
&{}+\theta^{\mu\nu}_{\hphantom{\mu\nu}a}T^{\rho}_{\mu\nu}\Tt^a_{\rho\lambda}\nonumber \\
&{}+\theta^{\mu\nu}_{\hphantom{\mu\nu}a}\left(\nabla_\lambda\Tt^a_{\mu\nu}\right.
\nonumber\\
&{}+\left.\nabla_\nu\Tt^a_{\lambda\mu}+\nabla_\mu\Tt^a_{\nu\lambda}\right)\nonumber \\
&{}+\tau^{\hphantom{\mathrm{B}}\mu}_{\mathrm{B}\hphantom{\mu}\nu}T^\nu_{\mu\lambda}- J^\mu
F_{\mu\lambda}.
\label{Belcons}
\end{align}
While this expression is quite unwieldy to say the least, it satisfies an important property: it reduces to
$\nabla_\mu\tau^{\hphantom{\mathrm{B}}\mu}_{\mathrm{B}\hphantom{\mu}\lambda}=- J^\mu
F_{\mu\lambda}$ when the {\em full} torsion $T^\lambda_{\mu\nu}=0$ throughout the spacetime region
in question. (To
see this, note that the projection of torsion to reduced torsion is linear, and the coefficients involve
the vielbeins, which are covariantly constant.) Compared with
the continuity equation Eq.\ (\ref{stresscont}), the spin current times curvature tensor term in that
equation has disappeared, though the terms $\theta\nabla \Tt$ are related to it,
in view of the second Bianchi identity\cite{Carroll2004}
\begin{equation}
\partial_{[\mu}
T^a_{\nu\lambda]}+\omega_{[\mu\hphantom{a}|b|}^{\hphantom{\mu}a}T^b_{\nu\lambda]}
=R_{[\mu\nu\hphantom{a}|b|}^{\hphantom{\mu\nu}a}e_{\lambda]}^b\label{antisymb}
\end{equation}
(recall that the indices surrounded by vertical bars $|\cdots|$ are {\em not} antisymmetrized with the
others, namely $\mu$, $\nu$, and $\lambda$).

Additionally, by considering invariance of a general action under internal spatial rotations, we find that
the Belinfante stress tensor satisfies the symmetry condition
\begin{equation}
\epsilon^a_{\hphantom{a}b}\tau^{\hphantom{\mathrm{B}}b}_{\mathrm{B}\hphantom{b}a}=
\epsilon^a_{\hphantom{a}b}\theta^{\mu\nu}_{\hphantom{\mu\nu}a}\Tt^b_{\mu\nu},\label{belinfantesym}
\end{equation}
so that it is symmetric in the absence of reduced torsion, even when the full torsion tensor is
nonvanishing. We thus see that our definition of $\tau_B$ reduces to the standard Belinfante
energy-momentum-stress tensor in the absence of torsion, and is symmetric in the presence of torsion
provided the reduced torsion vanishes. Therefore, we claim that it represents the physical stress
tensor of a general system.

An important special case that illustrates the significance of reduced torsion and the Belinfante
construction is a $2+1$-dimensional manifold with vielbeins that differ from the trivial ones
$e_\mu^\alpha=\delta_\mu^\alpha$ only in the space-space components $e_i^a$, and are independent of the
space coordinates.
If we try to directly solve the Cartan equations Eq.\ (\ref{cartan}) for the spin connection with the
torsion set to zero, we find that they are inconsistent. In fact, for this geometry
\begin{equation}
C^{ab}=\eta^{cb}e^i_c\partial_0e_i^a+\eta^{ac}e^i_c\partial_0e_i^b,
\end{equation}
and hence the spacelike torsion does not vanish for any choice of spin connection. The reduced torsion,
however, may be set to any arbitrary value, and the Cartan equations can be solved to find a spin
connection that is gauge equivalent to Eq.\ (\ref{spincon}). Note that this geometry is precisely what is
usually considered for computations of the viscosity tensor\cite{Avron1995,Read2009,Read2011,Bradlyn2012},
although the non-vanishing of the spacelike torsion has not previously been noted to our knowledge.
Our generalized Belinfante construction ensures the existence of a symmetric stress tensor provided one
takes the reduced torsion to be zero. This is done implicitly in the condensed matter literature, since, as
noted above, the reduced torsion does not enter into any usual microscopic actions.

There exists an explicit formula for the improvement term needed to
convert the canonical stress tensor into this Belinfante form. It can be derived from
\begin{equation}
\tau^{\hphantom{\mathrm{B}}\mu}_{\mathrm{B}\hphantom{\mu}\alpha}-\tau^\mu_{\hphantom{\mu}\alpha}
=-\frac{1}{\widehat{\sqrt{g}}}\int{d^{d+1}x\widehat{\sqrt{g}}J_{\mathrm{S}
\hphantom{\lambda}a}^{\hphantom{\mathrm{S}}\lambda\hphantom{a}b}\left(\frac{\delta
\omega_{\lambda\hphantom{a}b}^{\hphantom{\lambda}a}}{\delta e_\mu^\alpha}\right)_{\Tt^a}},
\end{equation}
although the general expressions are quite cumbersome and unilluminating. We note only that, in the absence
of torsion, the improvement term reduces to the known Belinfante improvement term, see for example Refs.\
\onlinecite{Martin,Weinberg}.

\section{Linear Response calculation of thermoelectric coefficients for non-interacting
electrons}\label{app}

In this appendix, we recapitulate the standard linear response calculation of the response functions
for non-interacting electrons in an integer quantum Hall phase.
\subsection{Operator Formalism}
We consider a model Hamiltonian for a system of electrons in a magnetic field
\begin{eqnarray}
H_0&=&\sum_p{\frac{\pi_i^p\pi_i^p}{2m}+V(\mathbf{r}_p)},\\
&=&\sum_p h^p
\end{eqnarray}
where we use $i,j=1,2$ for spatial indices as above, and $p,q=0,1\dots N$ for particle indices; $h^p$ is
the Hamiltonian for the $p$th particle. The
$\pi_i^p$ are the kinetic momenta, and
\begin{align}
\left[r_i^p,\pi_j^q\right]&=i\delta_{pq}\delta_{ij}, \\
\left[\pi_i^p\pi_j^q\right]&=iB\epsilon_{ij}\delta_{pq},
\end{align}
with $B$ the magnetic field strength. As our goal will be to calculate thermoelectric coefficients, in
particular the thermal conductivity tensor $\kappa_{ij}$, we need to identify the number current density
$J_i(\mathbf{r})$, the heat current density $J^\mathrm{Q}_i(\mathbf{r})$, and the perturbations to which
they couple.

Following Luttinger and CHR, we introduce a Hamiltonian density $h(\mathbf{r})$, a perturbing electric
field $\phi(\mathbf{r})$, and a fictitious gravitational field $\psi(\mathbf{r})$, and identify the
perturbed Hamiltonian as
\begin{align}
H_T&=\int d^2r\left[\left(1+\psi(\mathbf{r})\right)h(\mathbf{r})+\phi(\mathbf{r})\rho(\mathbf{r})\right]\\
&=\int d^2rh_T(\mathbf{r}),
\end{align}
where
\begin{equation}
\rho(\mathbf{r})=\sum_p\delta(\mathbf{r}-\mathbf{r}_i)
\end{equation}
is the density operator, and the Hamiltonian density $h(\mathbf{r})$ satisfies
\begin{equation}
\int{d^2rh(r)}=H_0.
\end{equation}
Note that there is an operator ordering ambiguity inherent in any attempt to define the energy density
$h(\mathbf{r})$. It will be essential later that we adopt the definition
\begin{equation}
h(\mathbf{r})=\sum_p\left(\frac{1}{2m}\pi_i^p\delta(\mathbf{r}-\mathbf{r}_i)\pi_i^p+V(\mathbf{r_i})
\delta(\mathbf{r}-\mathbf{r}_i)\right).
\end{equation}
This differs from the more commonly used expression $h^{\mathrm{CHR}}(\mathbf{r})$ by
\begin{equation}
h(\mathbf{r})=h^{CHR}(\mathbf{r})+\frac{1}{2m}\nabla^2\rho(\mathbf{r}),
\end{equation}
and instead corresponds closely with the second-quantized energy density operator of Section \ref{sec2}
used more recently in the literature\cite{Nomura2012,Sumiyoshi2012}. Note that $h(\mathbf{r})$ yields a
positive definite kinetic energy density, while $h^{CHR}(\mathbf{r})$ does not. This justifies its use,
contrary to established convention.

The number and energy currents are determined up to a divergence-free part by the continuity equations
\begin{align}
\frac{\partial\rho}{\partial t}+\nabla\cdot\mathbf{J}&=0, \label{numbercont}\\
\frac{\partial h_T}{\partial t}+\nabla\cdot\mathbf{J}^{\mathrm{E}}&=\left(\frac{\partial\phi}{\partial
t}+\frac{\partial\psi}{\partial t}\right)h. \label{energycont}
\end{align}
[the RHS of Eq.\ (\ref{energycont}) accounts for the fact that the \emph{explicit} time dependence of the
perturbing fields breaks energy conservation]. In order to fix the divergence-free pieces of the currents,
we demand that the CHR scaling relations
\begin{align}
\mathbf{J}(\mathbf{r})&=(1+\psi(\mathbf{r}))\mathbf{j}(\mathbf{r}) \label{numberscaling} \\
\mathbf{J}^{\mathrm{E}}(\mathbf{r})&=(1+2\psi(\mathbf{r}))\mathbf{j}^E(\mathbf{r})+\phi(\mathbf{r})\mathbf{j}
(\mathbf{r}) \label{energyscaling}
\end{align}
hold to first order in $\phi$ and $\psi$, where $\mathbf{j}(\mathbf{r})$ and $\mathbf{j}^{E}(\mathbf{r})$
are the unperturbed number and energy currents, respectively. These have exactly the form obtained from the
formalism in Sec.\ \ref{sec2}, with $e_\mu^0=\delta^0_\mu(1+\psi)$. A short calculation for the number
current reveals the standard result
\begin{equation}
j_i(\mathbf{r})=\frac{1}{2m}\sum_p\left\{\pi_i^p,\delta(\mathbf{r}-\mathbf{r}_i)\right\}.
\end{equation}
and for the energy current
\begin{align}
j_i^{E}(\mathbf{r})=&\frac{1}{2m}\sum_p\left\{\pi_i^p,\frac{1}{2m}\pi_j^p\delta(\mathbf{r}-\mathbf{r}_i)
\pi_j^p+
V(\mathbf{r}_i)\delta(\mathbf{r}-\mathbf{r}_i)\right\}\nonumber\\
&-\frac{i}{8m^2}\epsilon_{ij}\partial_j\left(\epsilon_{kl}\pi_k^p\delta(\mathbf{r}-\mathbf{r}_i)
\pi_l^p\right). \label{energycurrent}
\end{align}
We are now interested in the linear response of the total (or integrated) currents to the perturbations
$-\nabla\phi$ and $-\nabla\psi$, to lowest order in wavevector $\mathbf{q}$. Denote the integrated currents
by
\begin{align}
\bar{\mathbf{J}}&=\frac{1}{V}\int{d^2r\mathbf{J}(\mathbf{r})}=\frac{1}{mV}\sum_p\mathbf{\pi}^p,\\
\bar{\mathbf{J}}^\mathrm{E}&=\frac{1}{V}\int{d^2r\mathbf{J}^{\mathrm{E}}(\mathbf{r})}=\frac{1}{2mV}\sum_p
\left\{\mathbf{\pi}^p,h^p\right\},
\end{align}
where $V$ is the volume of the system. Since we are interested in vanishing $\mathbf{q}$, we may take for
the perturbations
\begin{align}
\phi(\mathbf{r})&=-E_i(t)r_i, \\
\psi(\mathbf{r})&=-G_i(t)r_i,
\end{align}
where we have restored the explicit time dependence of the perturbation. Then, what we want to compute are
the zero frequency response coefficients $R^{(n)}_{ij}$ satisfying
\begin{align}
\delta\left<\bar{J}_i\right>&=R^{(1)}_{ij}E_j+R^{(2)}_{ij}G_j, \label{linresponse1} \\
\delta\left<\bar{J}^{E}_{i}\right>&=R^{(3)}_{ij}E_j+R^{(4)}_{ij}G_j. \label{linresponse2}
\end{align}
Now, the response functions $R^{(n)}$ are not simply given by the naive
Kubo formulas; the scaling relations Eqs.\ (\ref{numberscaling}) and (\ref{energyscaling}) imply the
presence of contact terms. For $R^{(1)}_{ij}$ this is not the case and we have simply that
\begin{equation}
R^{(1)}_{ij}=\sigma_{ij}, \label{R1eqn}
\end{equation}
the usual zero-frequency conductivity tensor. For $R^{(2)}_{ij}$ we have in linear response
\begin{align}
R^{(2)}_{ij}&=L^{(2)}_{ij}-\frac{1}{V}\int{d^2r \left<r_{j}j_{i}(\mathbf{r})\right>_0}, \label{R2eqn}\\
L_{ij}^{(2)}&=i\int_0^{\infty}dt\,{e^{i\omega^+t}\left<\left[\bar{J}_i(t),D_j^{\mathrm{E}}(0)\right]\right>_0},
\end{align}
where $D_{i}^{\mathrm{E}}$ is the operator which couples to $G_i$ in the Hamiltonian, henceforth referred
to as the \emph{energy polarization}:
\begin{align}
D_{i}^{\mathrm{E}}&=\int{d^2r r_i h(\mathbf{r})} \\
&=\frac{1}{2}\sum_p\left\{h^p,r_i^p\right\}.
\end{align}
Defining the magnetization density $m_0$ to be
\begin{equation}
m_0=\frac{1}{2V}\int{d^2r\left<\mathbf{r}\times\mathbf{j}(\mathbf{r})\right>} \label{magdef2},
\end{equation}
we see, after some elementary manipulations, that the contact term in Eq.\ (\ref{R2eqn}) is simply
\begin{equation}
-\frac{1}{V}\int{d^2r \left<r_{j}j_{i}(\mathbf{r})\right>_0}=m_0\epsilon_{ij}.
\end{equation}
This term must be present due to the scaling relation (\ref{numberscaling}), although from the form above
we see that it is nonvanishing only when time-reversal symmetry is broken. Similarly, we have for
$R^{(3)}_{ij}$
\begin{align}
R^{(3)}_{ij}&=L^{(3)}_{ij}-\frac{1}{V}\int{d^2r \left<r_{j}j_{i}(\mathbf{r})\right>_0}, \label{R3eqn}\\
L_{ij}^{(3)}&=i\int_0^{\infty}dt\,{e^{i\omega^+t}\left<\left[\bar{J}^{\mathrm{E}}_i(t),D_j(0)\right]
\right>_0},
\end{align}
where $D_i$ is the ordinary polarization, which couples to $E_i$ in the Hamiltonian, i.e.,
\begin{equation}
D_i=\sum_pr_i^p.
\end{equation}
Note that the scaling relation (\ref{energyscaling}) ensures that it is the ordinary magnetization that
again appears as the contact term in Eq.\ (\ref{R3eqn}).

Up to now, these formulas have all agreed with those of Str\v eda and Smr\v cka. For $R^{(4)}$ however, we
have
\begin{align}
R^{(4)}_{ij}&=L^{(4)}_{ij}-\frac{1}{V}\int{d^2r \left<r_{j}j^{\mathrm{E}}_{i}(\mathbf{r})\right>_0},
\label{R4eqn}\\
L_{ij}^{(4)}&=i\int_0^{\infty}dt\,{e^{i\omega^+t}\left<\left[\bar{J}^{\mathrm{E}}_i(t),
D^{\mathrm{E}}_j(0)\right]
\right>_0},
\end{align}
he contact term can again be expressed in terms of a suitably defined energy magnetization
\begin{align}
2m_0^{\mathrm{E}}&=\frac{1}{V}\int{d^2r\left<\mathbf{r}\times\mathbf{j}^{\mathrm{E}}(\mathbf{r})\right>_0}
\\
&=\frac{1}{V}\sum_p\epsilon_{ij}\left<h^p\left\{r_i^p,\pi_{j}^p\right\}\right>_0-\frac{B\bar{n}}{4m^2}
\label{emag}
\end{align}
as
\begin{equation}
-\frac{1}{V}\int{d^2r \left<r_{j}j^{\mathrm{E}}_{i}(\mathbf{r})\right>_0}
=(2m_0^{\mathrm{E}}+\frac{B\bar{n}}{4m^2})\epsilon_{ij}.
\end{equation}
It is important to note that in deriving these contact terms, and in relating them to the magnetizations,
certain integrals and trace identities need to be used which are only valid if the states live in an
honest-to-goodness Hilbert space---i.e.\ if they are normalizable. Thus, the presence of the confining
potential $V(\mathbf{r})$ is indispensable at this stage of the calculation. It is only in the final
expressions in Subsection \ref{asec4} where we will be able to take $V\rightarrow 0$.

It is also worth mentioning that these Kubo formulas can be put into a different, more suggestive form.
Using the identities for the integrated currents
\begin{align}
\bar{J}_{i}&=\frac{1}{V}\frac{\partial D_i}{\partial t}, \\
\bar{J}_{i}^{\mathrm{E}}&=\frac{1}{V}\frac{\partial D_i^{\mathrm{E}}}{\partial t},
\end{align}
we can integrate Eqs. (\ref{R1eqn},\ref{R2eqn},\ref{R3eqn},\ref{R4eqn}) by parts to find
\begin{align}
R^{(1)}_{ij}&=\frac{\omega^+}{V}\int_0^{\infty}{dte^{i\omega^+t}\left<\left[D_i(t),D_j(0)\right]\right>},
\label{edge1} \\
R^{(2)}_{ij}&=\frac{\omega^+}{V}\int_0^{\infty}{dte^{i\omega^+t}\left<\left[D^{\mathrm{E}}_i(t),D_j(0)\right]
\right>},\label{edge2} \\
R^{(3)}_{ij}&=\frac{\omega^+}{V}\int_0^{\infty}{dte^{i\omega^+t}\left<\left[D_i(t),D^{\mathrm{E}}_j(0)
\right]\right>},\label{edge3} \\
R^{(4)}_{ij}&=\frac{\omega^+}{V}\int_0^{\infty}{dte^{i\omega^+t}\left<\left[D^{\mathrm{E}}_i(t),
D^{\mathrm{E}}_j(0)\right]\right>},\label{edge4}
\end{align}
where the surface terms arising from the partial integration exactly cancel the magnetization contact terms
(in the case of the conductivity $R^{(1)}$, both are identically zero).
In this form, we know from the projection theorem of Ref.\ \onlinecite{Bradlyn2012}, that in the
thermodynamic limit as $\omega\rightarrow 0$, the $R^{(n)}_{ij}$ will be dominated (if there were no
confining potential) by matrix elements of the polarization operators coming from states degenerate with
the ground state. In the presence of the confining potential, however, the center-of-mass degeneracy of the
Landau-levels is broken, and there is no longer an exact degeneracy. On the other hand, in the
thermodynamic limit, edge excitations become gapless, and in fact have a linear dispersion. The sum over
matrix elements then, schematically, produces terms like
\begin{equation}
\omega^+\int{d\mathbf{k}\rho(\mathbf{k})\frac{F_{0\mathbf{k}}}{\omega^+-v\mathbf{k}}}
\end{equation}
where $\rho(\mathbf{k})$ is the density of states for the edge excitations. The functions
$F_{0\mathbf{k}}$ represent the matrix elements of the polarization operators between the ground state
and the various edge-excited states. These can be interpreted as moments of the energy density operator
on the edge (since in the bulk the states are indistinguishable). This integral, when viewed as a function
of $\omega$, has a branch point at the origin, and therefore the limit $\omega\rightarrow 0$ must
be evaluated carefully: it will be non-vanishing.

This demonstrates clearly the role of edge states in determining the thermoelectric coefficients. One
must keep in mind, however, that the conductivity $R^{(1)}$ is fairly special in this regard. In the
absence of a confining potential, the polarization-polarization and current-polarization response
functions are completely equivalent; the contributions to the conductivity can be viewed alternatively
as coming from the center-of-mass degenerate single-particle states in the thermodynamic limit. For the
other response functions, the current-polarization form must be added to the magnetization contribution
in order to recover the full response function. In the absence of a confining potential, the magnetization
term is not well defined: particles at larger and larger distances contribute more and more to the
magnetization. Thus, for these response function, the presence of edge states is essential.

\subsection{Zero Temperature Response in the IQH Regime}\label{asec3}

We would now like to explicitly calculate the $R^{(n)}$'s for a system with $j$ filled Landau levels
at chemical potential $\mu\in\left[\omega_c(j-1/2),\omega_c(j+1/2)\right]$ lying in the bulk gap between
levels. We are interested primarily in the low-temperature behavior of the response coefficients. However,
for single-particle operators we have
\begin{align}
\left<\mathcal{O}\right>(\mu,T)&=\int_{-\infty}^{\infty}d\eta \,
n_\mathrm{F}(\eta,T)\tr(\delta(\eta-h)\mathcal{O}) \\
&=-\int_{-\infty}^{\infty}d\eta
\frac{dn_\mathrm{F}}{d\eta}(\eta,T)\int_{-\infty}^{\eta}d\zeta\tr(\delta(\zeta-h)\mathcal{O}) \\
&=-\int_{-\infty}^{\infty}d\eta \frac{dn_\mathrm{F}}{d\eta}(\eta,T)\left<\mathcal{O}\right>(\mu=\eta,T=0),
\label{finitetempreln}
\end{align}
where the trace is over all single-particle states, $h$ is the single-particle Hamiltonian, and
$n_F(\eta,T)=1/(e^{(\eta-\mu)/T}+1)$ is the Fermi function. Thus, we can determine the behavior of the
response coefficients at nonzero temperature once their zero-temperature behavior is known. Hence, in this
section we will aim to evaluate the $R^{(n)}_{ij}$ at zero temperature. This was done in Ref.\
\onlinecite{Smrcka1977} using a resolvent formalism, however here we will proceed directly using the Kubo
formulas in the time domain. This will allow us to illuminate some subtleties in the derivation. For
notational simplicity, it should be understood that the limit $\omega\rightarrow 0$ is implied in all
expressions in this section

Let us begin by noting that the conductivity $\sigma_{ij}$ is given by
\begin{align}
\sigma_{ij}(\mu)&=\frac{i}{mV}\int_{-\infty}^\mu
d\eta\int_0^{\infty}dte^{i\omega^+t}\tr(\delta(\eta-h)\left[\pi_i(t),r_j(0)\right])\\
&\equiv\int_{-\infty}^{\mu}d\eta A_{ij}(\eta), \label{stredaA}
\end{align}
where we have used the freedom afforded us by this noninteracting problem to evaluate the averages using
only single-particle states. As its name would suggest, this $A_{ij}$ is a generalization of the function
introduced by Smr\v cka and Str\v eda\cite{Smrcka1977}.

Next, we examine $R^{(3)}_{ij}$ ($R^{(3)}=R^{(2)}$ via Onsager
reciprocity\cite{Cooper1997,Smrcka1977}). We can write the Kubo part of the response function,
$L^{(3)}_{ij}$, as
\begin{align}
L_{ij}^{(3)}
&=\int_{-\infty}^{\mu}{d\eta\left( \eta A_{ij}(\eta)+\frac{1}{2}B_{ij}(\eta)\right)},
\end{align}
where we have defined
\begin{equation}
B_{ij}(\eta)=\frac{1}{m^2V}\int_0^{\infty}dte^{i\omega^+t}\tr(\delta(\eta-h)\left\{\pi_i(t),\pi_j(0)
\right\}). \label{stredaB}
\end{equation}
After a partial integration, we find that
\begin{equation}
L^{(3)}_{ij}=\mu\sigma_{ij}(\mu)+\int_0^{\mu}d\eta\left(\eta-\mu\right)\left(A_{ij}-\frac{1}{2}\frac{dB_{ij}}
{d\eta}\right). \label{L3almost}
\end{equation}
We can perform a similar analysis for $L^{(4)}_{ij}$ to find
\begin{align}
L_{ij}^{(4)}
&=\frac{B\bar{n}}{4m^2}\epsilon_{ij}\nonumber \\
&+\mu^2\sigma_{ij}(\mu)+\int_{-\infty}^{\mu}{d\eta\left(\eta^2-\mu^2\right)\left(A_{ij}(\eta)-\frac{1}{2}
\frac{dB_{ij}}{d\eta}\right)}. \label{L4almost}
\end{align}

Now, it can be shown\cite{Smrcka1977} that
\begin{equation}
A_{ij}-\frac{1}{2}\frac{dB_{ij}}{d\eta}=\epsilon_{ij}\frac{dm_0}{d\eta}. \label{stredarelation}
\end{equation}
Plugging this into Equations (\ref{L3almost}) and (\ref{L4almost}), we find
\begin{align}
L^{(3)}_{ij}&=\mu\sigma_{ij}(\mu)-m_0\epsilon_{ij}, \label{L3final}\\
L^{(4)}_{ij}&=\mu^2\sigma_{ij}(\mu)-2m_0^\mathrm{E}\epsilon_{ij},\label{L4final}
\end{align}
where we have used the relation
\begin{equation}
m^\mathrm{E}=\int_{-\infty}^{\mu}d\eta \eta m_0(\eta)-\frac{B\bar{n}}{8m^2},
\end{equation}
which follows from Eq.\ (\ref{emag}). From our discussion above, we recognize the magnetization terms in
Eqs.\ (\ref{L3final}--\ref{L4final}) as precisely the negative of the contact terms in Eqs.\ (\ref{R3eqn})
and (\ref{R4eqn}). Furthermore, we see that the explicit dependence on the chemical potential in the first
terms above indicates that these are edge contributions. Upon subtracting the total magnetizations, we
have that the $L$'s are given by the \emph{bulk} contributions to the magnetizations, as asserted in Sec.\
\ref{respsec} and consistent with Eqs.\ (\ref{l1}-\ref{l4}).  Thus, putting everything together, we have
\begin{align}
R^{(1)}_{ij}(\mu)&=\sigma_{ij}(\mu), \\
R^{(2)}_{ij}(\mu)&=R^{(3)}_{ij}(\mu)=\mu\sigma_{ij}(\mu), \\
R^{(4)}_{ij}(\mu)&=\mu^2\sigma_{ij}(\mu),
\end{align}
in agreement with known results.

\subsection{Extension to Nonzero Temperature} \label{asec4}

Having derived expressions for the $R^{(n)}$ at zero temperature, we can now use Eq.\
(\ref{finitetempreln}) to evaluate the transport coefficients for all values of chemical potential $\mu$
and temperature $T$. Let us start with the Hall conductivity $R^{(1)}$. As is well known, at zero
temperature we have in the thermodynamic limit (this is the stage at which it is safe to take the limit)
and with chemical potential $\mu$ in a bulk gap
\begin{equation}
R^{(1)}_{ij}(\mu)=\frac{1}{2\pi}\epsilon_{ij}\sum_n\Theta(\mu-\epsilon_n),
\end{equation}
where $n$ indexes the Laudau levels, and $\epsilon_n$ is the Landau level energy. At nonzero temperature,
this becomes
\begin{align}
R^{(1)}_{ij}(\mu,T)&=-\int_{-\infty}^{\infty}{d\eta\frac{dn_F}{d\eta}R^{(1)}_{ij}(\eta,0)}\nonumber \\
&=-\epsilon_{ij}\frac{1}{2\pi}\sum_n\int_{-\infty}^{\infty}{d\eta
\frac{dn_F(\eta)}{d\eta}\Theta(\eta-\epsilon_n)} \\
&=\frac{1}{2\pi}\sum_nn_F(\epsilon_n)\epsilon_{ij},
\end{align}
from which we see that corrections to the low temperature behavior are exponentially suppressed, as
expected. In fact, we have for $\eta$ in a neighborhood of the chemical potential $\mu$, that
$R^{(1)}(\eta)$ is a slowly varying function when $\mu$ is in a gap (actually, it is a constant), and
hence for temperatures $T\ll\omega_c$ we can make use of the Sommerfeld expansion
\begin{equation}
-\frac{dn_F}{d\eta}\approx\delta(\mu-\eta)+\frac{\pi^2}{6}T^2\delta''(\mu-\eta)+\dots,
\end{equation}
from whence we see that the corrections to the conductivity at low temperature are non-perturbatively
suppressed.

Using similar logic, we find for the thermoelectric transport
coefficients\cite{Streda1982,Streda1983,Jonson1982,Jonson1984}
\begin{align}
N^{(1)}_{ij}(\mu,T)&=\frac{\nu}{2\pi}\epsilon_{ij}, \\
N^{(2)}_{ij}(\mu,T)&=0, \\
N^{(3)}_{ij}(\mu,T)&=0, \\
N^{(4)}(\mu,T)&=\frac{\pi\nu T^2}{6}\epsilon_{ij},
\end{align}
As emphasized throughout, the non-zero contribution to $N^{(4)}$ is purely an edge effect.
We have for an integer quantum Hall system at low temperatures
\begin{align}
\bar{J}_{i}&=-\frac{\nu}{2\pi}\epsilon_{ij}\partial_{j}\phi, \\
\bar{J}_{i}^\mathrm{Q}&=-\frac{\pi\nu T}{6}\epsilon_{ij}\partial_{j}T.
\end{align}
We see directly from this that the thermal Hall conductivity is given by Eq.\ (\ref{thermalhallresult}),
with central charge $c=\nu$ corresponding to $\nu$ filled Landau levels. Note also the Wiedemann-Franz
relation
\begin{equation}
\kappa_{ij}=\frac{\pi^2 T}{3}\sigma_{ij}.
\end{equation}

\bibliography{eff-action-improved}
\end{document}